%% file: main.tex
\documentclass[conference,compsoc]{IEEEtran}

\ifCLASSOPTIONcompsoc
  \usepackage[nocompress]{cite}
\else
  \usepackage{cite}
\fi

\input{imports}

\usepackage{svg}
\usepackage{amssymb}
\usepackage{mdframed}
\usepackage[hidelinks]{hyperref}

\usepackage[available,functional]{ieeebadges}

\providecommand{\shortsection}[1]{\vspace{0.6ex}\noindent\textbf{#1.}\ }

\newcommand{\oldtxt}[1]{}

\begin{document}

\title{It's a Feature, Not a Bug: Secure and Auditable State\\Rollback for Confidential Cloud Applications}


%
\IEEEoverridecommandlockouts 
\author{%
\IEEEauthorblockN{
Quinn Burke\IEEEauthorrefmark{1}\thanks{\IEEEauthorrefmark{1}Corresponding author (email: qkb@cs.wisc.edu)}, Anjo Vahldiek-Oberwagner\IEEEauthorrefmark{2}, Michael Swift, Patrick McDaniel}
\IEEEauthorblockA{University of Wisconsin-Madison, Intel Labs\IEEEauthorrefmark{2}}
}


\maketitle

\thispagestyle{plain}
\pagestyle{plain}

\begin{abstract}
\input{00}
\end{abstract}


%
\IEEEpeerreviewmaketitle
\input{01}
\input{02}
\input{03}
\input{04}
\input{05}
\input{06}
\input{07}
\input{08}
\input{acks}

\appendices



\bibliographystyle{IEEEtran}
\bibliography{IEEEabrv,main}

\input{appx-1}

\input{meta_review}

\end{document}

%% file: imports.tex
\usepackage{amsthm}                              
\usepackage{amsthm}                              
\usepackage[ruled,linesnumbered]{algorithm2e}    
\usepackage{algpseudocode}
\SetCommentSty{itshape}
\usepackage{booktabs}                            
\usepackage{centernot}                           
\usepackage{colortbl}                            
\usepackage{graphicx}                            
\usepackage{svg}                                 
\usepackage[T1]{fontenc}                         
\usepackage{hhline}                              
\usepackage{lipsum}                              
\usepackage{mathtools}                           
\usepackage[all]{nowidow}                        
\usepackage{siunitx}                             
\usepackage{substr}                              
\usepackage{tikz}                                
\usetikzlibrary{arrows.meta,positioning,shapes,fit,calc} 
\usepackage{tcolorbox}                           
\usepackage{xcolor}                              
\usepackage{circledsteps}
\usepackage{xspace}                              
\usepackage{multirow}
\usepackage{float}
\usepackage{url}
\usepackage{pgf}                                
\usepackage{wrapfig}
\usepackage{pifont}
\usepackage{epsfig}
\usepackage{endnotes}
\usepackage{amssymb}
\usepackage{amsfonts}
\usepackage{textcomp}
\usepackage{comment}
\usepackage{enumitem} 
\usepackage{microtype}
\usepackage{tabularx}
\usepackage{threeparttable}                      
\usepackage{wasysym}                             

\usepackage{amsthm}                               

\newcommand{\tool}{\textsc{Rebound}}

\newcommand{\red}[1]{\textcolor{black}{#1}}


%% file: 00.tex
Replay and rollback attacks threaten cloud application integrity by reintroducing authentic yet stale data through an untrusted storage interface to compromise application decision-making. Prior security frameworks mitigate these attacks by enforcing forward-only state transitions (state continuity) with hardware-backed mechanisms, but they categorically treat all rollback as malicious and thus preclude legitimate rollbacks used for operational recovery from corruption or misconfiguration. We present \tool{}, a general-purpose security framework that preserves rollback protection while enabling policy-authorized legitimate rollbacks of application binaries, configuration, and data. Key to \tool{} is a reference monitor that mediates state transitions, enforces authorization policy, guarantees atomicity of state updates and rollbacks, and emits a tamper-evident log that provides transparency to applications and auditors. We analyze \tool{}'s security properties and show through an application case study---with software deployment workflows in GitLab CI---that it enables robust control over binary, configuration, and raw data versioning with low end-to-end overhead.

%% file: 01.tex
\section{Introduction}
\label{sec:introduction}

Ensuring the integrity and freshness of storage is a fundamental challenge in cloud security. Even with trusted execution environments (TEEs), this is challenging because persistent state resides outside the TEE and may outlive the application, exposing it to a wide range of threats~\cite{tsai_graphene-sgx_nodate,galanou2023trustworthy}. While cryptographic mechanisms such as authenticated encryption using hardware-rooted keys address many security concerns for persistent state, they do not mitigate \textit{replay and rollback attacks}, where attackers with control of the storage interface reintroduce authentic yet stale data into the system to compromise application decision-making~\cite{johannesmeyer2024practical,Hoven2019}. 

To address these risks, a central goal of recent research has been on designing security frameworks that provide rollback protection by enforcing \textit{state continuity}. This property ensures all state transitions represent forward progress, and stale data cannot be reintroduced to the application without detection~\cite{parno2011memoir,angel2023nimble,matetic2017rote,strackx2016ariadne}. Note that a state transition can be represented broadly, from flushing a set of file modifications to disk, to deploying a new application binary to production servers. Typically, each state update is bound to a trusted, hardware-based \textit{monotonic counter}. State updates trigger counter increments, and all read data is cryptographically verified against the current counter value. Any state read by the application that does not reflect strictly forward progress (i.e., stale data) is rejected as malicious.

While this mitigates rollback \textit{attacks}, there are legitimate operational reasons to reintroduce earlier state to the system. Consider modern software deployment pipelines (e.g., GitLab CI, GitHub Actions, Tekton), which continuously build and deploy versioned bundles of artifacts containing container image digests, deployment and service manifests, and configuration and secret material, among other objects~\cite{alluri2020devops,zampetti2021ci}. Administrators are often forced to revert to prior releases when a newly deployed version crashes, becomes misconfigured, or exposes a latent vulnerability~\cite{marques2023foundational}. Such \textit{authorized rollbacks} are a core DevOps resilience primitive, allowing administrators to quickly restore service health by reinstating previous releases. Yet, prior state continuity frameworks provide no means to securely support this operational requirement---their threat model assumes \textit{all} rollbacks are adversarial. What is needed is a unified framework that guarantees rollback protection while permitting authorized rollbacks in a secure and auditable manner. 

In this paper, we develop a new general-purpose security framework \tool{} that enables secure rollback while preserving state continuity guarantees. \tool{} can be instantiated as a standalone service. It can interpose on (and protect) all state-changing operations made by applications, such as software releases, application writes, etc. Rather than enforcing forward-only state transitions, \tool{} extends prior works by permitting selective and arbitrary-point rollback to historical states, while preserving state continuity. In \tool{}, rollback events are explicitly declared, authorized against policy, recorded in a tamper-evident log, and sealed into future state. \tool{} consists of a reference monitor that mediates all state transitions, tracks and authenticates historical state, ensures atomicity of state transitions, and provides fine-grained auditability to benign clients, administrators, or third-parties.

Designing such a framework requires addressing several new challenges. First, simultaneously providing rollback protection and authorized rollbacks requires new mechanisms for \textit{managing, authenticating, and reviving historical state} while preserving state continuity guarantees. We address this through new abstractions and data structures that prevent attackers from tampering with any step in a state update or rollback operation. Second, rollback and recovery operations must be \textit{atomic and crash-consistent}, preventing adversaries from exploiting system crashes to stage unauthorized state transitions. We address this through new atomic commit and recovery protocols. Third, detailed audit trails are desirable and routinely required by compliance regulations for high-assurance applications (financial, medical, etc.). All state transitions must thus be \textit{observable and accountable} (i.e., attributable to initiators, causes, etc.). We address this by designing a reference monitor that governs all state transitions and manages a tamper-evident audit log that records all state updates, rollback events, policies, and justifications in a verifiable, queryable format. Finally, all of the above challenges must be addressed \textit{efficiently and at scale} to meet demands of modern cloud applications. We address this by introducing additional control knobs to tune performance and storage overhead (e.g., state pruning at a threshold age) while preserving state continuity guarantees.

We evaluate \tool{} in two stages. Our security analysis \red{shows} that \tool{} prevents rollback attacks, can securely recover from adversarial system crashes, and provides a fine-grained, tamper-evident audit trail of all state transitions for auditors. We then implemented \tool{} as a library that can be linked to applications and protect arbitrary application- or storage-layer data objects. We evaluated it across a suite of micro-benchmarks and macro-benchmarks; the latter instantiated \tool{} as a standalone service that interposed on GitLab CI pipelines (i.e., \tool{} APIs were invoked during job execution to protect the pipeline artifacts). We showed that \tool{} delivers these capabilities with low end-to-end overhead: \tool{}'s state updates and rollbacks add <1 second of overhead for simple CI builds and $<5\%$ overhead for complex ones which can take several minutes (or hours) to complete. Overheads scale sublinearly as the number of releases grows, completing in \red{0.5--1} seconds per object even after 100 releases. Query latency scales logarithmically (<13ms after 100 releases), enabling real-time verification, and storage costs can be bounded via secure pruning while maintaining full cryptographic accountability.

Our work bridges the gap between rollback functionality expected by modern cloud applications and security assumptions made by prior security frameworks. \tool{} shows that rollbacks need not be treated as inherently malicious operations, but can be supported as a legitimate operation in a controlled and secure manner. \tool{} opens the door to a new class of secure applications that can take advantage of robust, hardware-rooted data version controls. The code is open-sourced at \red{\url{https://doi.org/10.5281/zenodo.19009015}}.

%% file: 02.tex
\section{Background}
\label{sec:background}

\subsection{Security Model}
\label{sec:threat-model}

\begin{figure}[t]
    \centering
    \includegraphics[width=\linewidth]{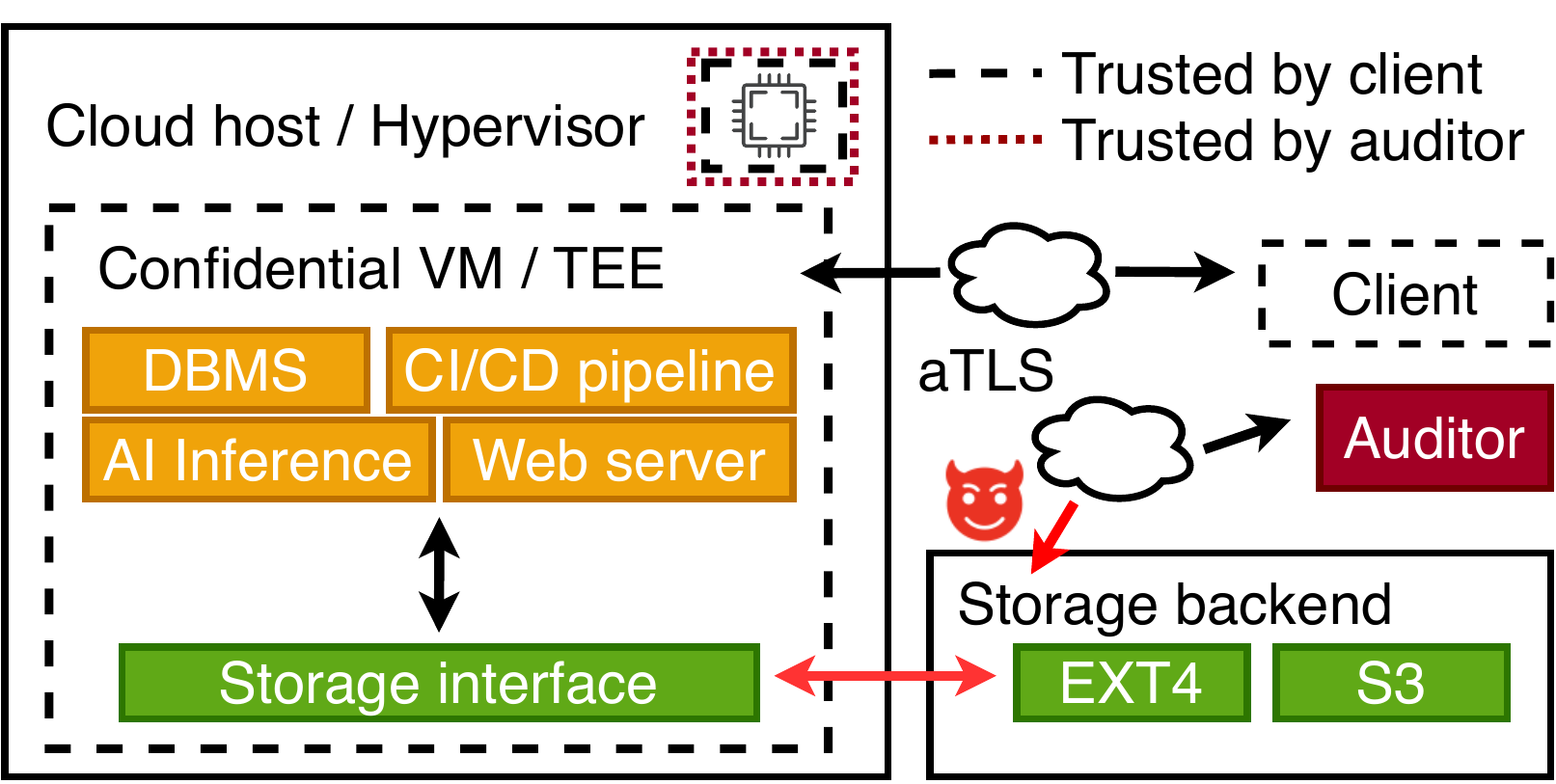}
    \caption{Confidential applications run inside TEEs on untrusted cloud infrastructure. The TEE protects code and data in memory and must check data integrity and freshness when reading/writing to storage.}
    \label{fig:tee-background}
\end{figure}

\shortsection{System Model} 
As shown in~\autoref{fig:tee-background}, we consider a deployment model where applications run inside trusted execution environments (TEEs) on public cloud infrastructure and read and write data to an untrusted storage system. These are commonly referred to as \textit{confidential cloud applications}. The storage interface can be file, object, or block, and the storage system can be either local or remote. Without loss of generality, we focus on a logically centralized storage interface. The physical distribution of the underlying storage (or application itself) is an orthogonal concern.

\shortsection{Trust Model} 
We assume there are three parties in the system: clients, cloud provider, and third-party auditors. Clients are operators, administrators, or other applications. Clients assume all application code and data in memory at the server are secure and trusted. This is guaranteed by memory isolation and remote attestation capabilities of commercial TEEs (Intel SGX, TDX, AMD SEV-SNP, etc.)~\cite{tsai_graphene-sgx_nodate,galanou2023trustworthy}. Clients therefore trust the platform hardware and running server code (TEE). Auditors also trust the platform hardware and TEE (after attestation). Initial versions of Intel SGX hardware included trusted counters whereas most recent TEEs (Intel TDX, AMD SEV) do not. Depending on the used hardware technology we assume it to be combined with a trusted platform module (TPM)~\cite{shang2024ccxtrust}, HSM~\cite{thales2025monotonic} or the availability of a secure monotonic counter service~\cite{martin2021mcs, fernandez2024vmcaas}.

\shortsection{Threat Model} 
We consider an \textit{external} attacker who controls the hypervisor and storage. This could be a malicious cloud administrator or a co-tenant with escalated privileges~\cite{parno2011memoir,angel2023nimble,matetic2017rote,strackx2016ariadne}. The attacker can tamper with any data crossing the storage interface. We also assume the attacker can control vCPU scheduling and arbitrarily delay program execution or crash the system to induce rollback attacks (i.e., lose buffered updates and force reversion to a prior state). Note that authenticated encryption protects against some forms of tampering (e.g., corruption), but not replay or rollback attacks. Lastly, we consider an authorized \textit{internal} attacker who has legitimate application access and can issue state-changing commands (e.g., rollback). This could be a compromised \red{client or administrator}. We will develop defenses against both types of attackers.

\subsection{State Continuity for Rollback Protection}
\label{sec:state-continuity}

Replay and rollback attacks threaten cloud application integrity by reintroducing authentic yet stale state into the system to corrupt application decision-making. The consequences range from an application consuming stale user data to a cluster controller reloading an older, vulnerable version of a binary. Prior works have addressed these attacks by introducing security frameworks that enforce \textit{state continuity}. These frameworks guarantee that state transitions advance only in forward order, ensuring each persisted update reflects the most recent authorized state. 

\begin{figure}[t]
    \centering
    \includegraphics[width=0.45\textwidth]{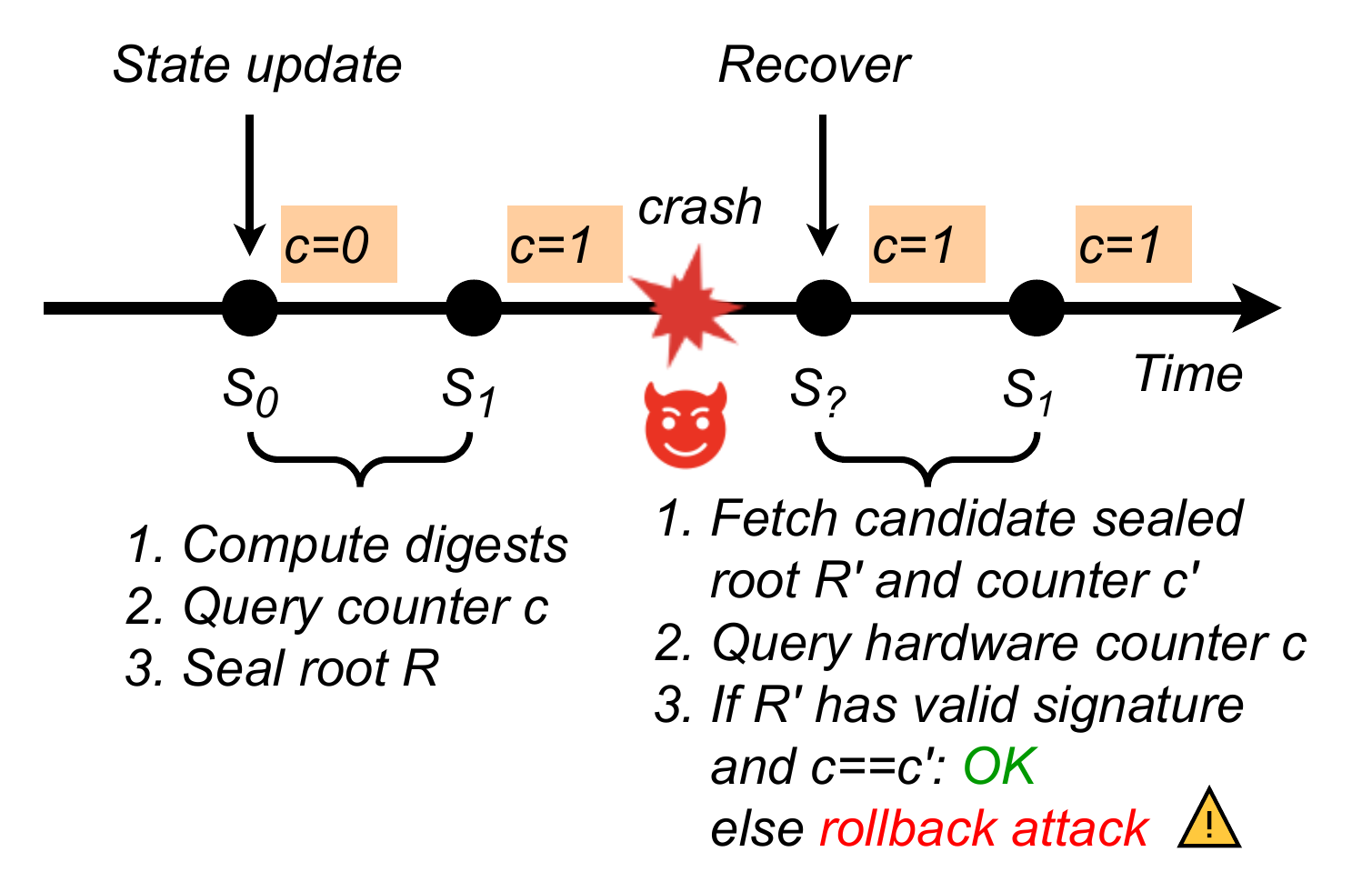}
    \caption{Conventional state continuity frameworks prevent rollback attacks by sealing state updates via tamper-evident mechanisms. However, they are not \textit{rollback-aware}: they only serialize opaque forward writes and cannot represent the intent of nor execute an authorized rollback.}
    \label{fig:continuity}
\end{figure}

Systems like Memoir~\cite{parno2011memoir}, ROTE~\cite{matetic2017rote}, Ariadne~\cite{strackx2016ariadne}, and Nimble~\cite{angel2023nimble} implement this in several steps (see Figure~\ref{fig:continuity}). First, all objects that compose the state are aggregated into a single state digest. This means that a digest of each object (e.g., a flat SHA256 hash) is computed, then aggregated into a global digest by further hashing the collection of per-object digests. The current monotonic counter value is then retrieved from the processor, concatenated with the global digest, and the result is hashed and signed. The signature represents the commit of the state change: the authoritative representation of the current system state. It can be saved to untrusted storage, and it is commonly referred to as the cryptographic \textit{seal}.

All application writes are required to invoke this new state update API. On future reads, the application first fetches a candidate sealed state from storage. It then verifies the signature, checks that the embedded counter is equivalent to the current counter value to ensure freshness, then authenticates read data against the embedded global digest. 

The security guarantees stem from hardware mechanisms preventing any entity from decrementing the counter. However, a consequence of this simple procedure is that there is no prescribed way for administrators to authorize, represent, or audit legitimate rollbacks.

\subsection{Rollback as a System-Level Feature}
\label{sec:feature-rollback}

Authorized rollback is a widespread feature across storage systems, databases, and deployment tooling. Modern file systems (ZFS, Btrfs~\cite{both2023btrfs,zhang_end--end_nodate}), databases (transactional abort, point-in-time recovery~\cite{yang2006trap,elnozahy2004checkpointing}), and software deployment pipelines~\cite{sillito2020failures,kerzazi2016botched} all provide mechanisms to revert state to prior versions. We focus particularly on software deployment pipelines such as GitLab CI as a running example, though our observations generalize to other domains.

\shortsection{Running Example: Software Deployment Pipelines}
Software deployment pipelines automate building, testing, and releasing software, managing multiple artifacts that must be versioned and deployed together: container images, configuration files, database schemas, and secrets.

A GitLab CI \textit{pipeline} specifies ordered \emph{stages} (e.g., build, test, deploy) containing \emph{jobs}. A successful promotion to \textit{production} must satisfy policy gates such as protected branch rules, required approvals, and vulnerability checks. The final output is a \emph{release bundle}: a logical state that includes container image digests, deployment manifests, environment configurations, and database migration markers. Deploying this bundle results in numerous low-level read and write operations that update image tags, configuration entries, and other state elements to reflect the new release.

When a bug is detected post-deployment, an administrator can initiate a rollback. The system must locate a prior release bundle and restore the state of each of its constituent artifacts. This process similarly results in numerous low-level read and write operations to revert image tags, configuration entries, and other state to their previous versions. Importantly, these systems implicitly assume storage is trusted, and this assumption does not hold in adversarial settings such as public clouds. We examine the implications of this below.

%% file: 03.tex
\section{\tool{} Overview}
\label{sec:design} 
This section describes \tool{}'s security properties, design goals, components, and core APIs.

\subsection{Security Properties \& Design Goals}
\label{sec:goals}
We define a secure rollback system as having the following security properties, which borrow from and extend those introduced in prior work~\cite{strackx2016ariadne}:

\begin{description}
\item[P1: State Continuity.] Unauthorized replay/rollback of persistent state should be prevented: every read should return fresh data or metadata. 

\item[P2: Secure Recovery.] All state transitions should be executed atomically (via transactions), and during recovery, attempts to replay in-flight (but not committed via a counter increment) transaction data should be prevented.

\item[P3: Observability \& Accountability.] State changes must be observed and mediated by a reference monitor, such that attribution (who/why) is bound at commit to the exact effects (what/when) in a single authoritative order. This enables verifiable answers about event completeness, exclusivity, and absence. 
\end{description}

Note that like prior state continuity frameworks, applications must be modified slightly (on the commit path, e.g., deploy or fsync) to invoke \tool{}'s APIs; compatibility is a non-goal. Our functional design goals are as follows:

\begin{description}
\item[G1: Authorized Rollback.] Rollback should be a first-class operation to authorized clients to facilitate rapid recovery from failed deployments or misconfigurations.

\item[G2: De-authorization \& Compliance.] De-authorization should be supported via pruning to mark versions as ineligible for rollback, enabling compliance with security policies (e.g., data retention windows) and removal of compromised versions.

\item[G3: Efficiency \& Scalability.] The security mechanisms (for P1, P2, and P3) should incur minimal overhead to application critical paths. For auditors, query latencies should be kept practical: logarithmic metadata growth and low query latency. Pruning (and thus de-authorization) should also be supported to bound storage costs of historical data.
\end{description}

\subsection{Straw-man Solutions and the Semantic Gap}
\label{sec:semantic-gap}

Rollbacks have been the subject of a vast amount of prior research in both the security and systems communities. A natural question is whether prior works already meet (or could be composed to meet) the properties and goals outlined above. This section briefly examines the straw-man approach and highlights vulnerabilities that emerge from simple applications of prior work.

Consider a straw-man design where an application runs on top of a conventional state continuity framework. Suppose the application maintains an audit log that records some information about state changes, such as timestamps and perhaps other metadata. Now consider that there exists a rollback utility or API that allows an administrator to revert any application state (binaries, configuration, or data) to a prior version. In this design, the application layer manages the audit log, while the state continuity layer underneath ensures rollback protection for storage. This separation of concerns creates two classes of vulnerabilities.

\shortsection{Lack of a Fresh Authentication Anchor}
The first issue arises because conventional continuity frameworks prescribe protocols that seal only the ``current bytes'' of an application. This means the \textit{latest} seal (i.e., the seal created under the current monotonic counter value) contains no authenticated commitments to history or policy decisions. Suppose an administrator wants to perform a rollback after detecting an issue with the deployed application. To authorize a rollback, the administrator must consult historical metadata (e.g., a snapshot registry) that resides on a separate, untrusted storage system. They must fetch the listing, select a tag, and verify it (along with the snapshot content) against its historical seal.

However, this historical seal is entirely disconnected from the current system state---there is no cryptographic binding between it and the latest seal. This violates state continuity (P1), as stale metadata must be accepted into the rollback decision process. In this scenario, an attacker controlling storage can perform a \textit{second-order rollback attack} (a type of second-preimage attack) by returning an old listing together with its matching historical seal and snapshot content. This would allow them to deceive the administrator into selecting an outdated, potentially vulnerable target of their choosing. This problem is amplified by lifecycle policies like snapshot pruning. The goal of pruning is to mark certain targets as ineligible. However, in this scenario, marking a version as ineligible is not reflected in the continuity layer's core logic because only current bytes are sealed. During rollback, the framework will thus validate historical content under its old seal, while remaining blind to the policy decision that now de-authorizes it. An attacker can similarly exploit this to maliciously resurrect pruned versions.

\shortsection{Loss of Accountability}
The more nuanced issue is a loss of accountability stemming from our trust model. Modern security standards and regulations like SOX and PCI DSS~\footnote{\red{\tool{} provides the foundational technical layer for such requirements (a tamper-evident, cryptographically bound record of all state changes) but full compliance also involves organizational controls (e.g., access review, separation of duties) that are outside the scope of this work.}} are driven by the principle of independent verification: the subject of an audit (the application) cannot also be the trusted source of evidence for that audit. In our setting, this means auditors do not trust the application or its logs; instead, they must rely on a separate attestable continuity layer (reference monitor) that mediates state changes and runs independently of the application code. An auditor therefore assumes that the application or its logs could be buggy or malicious. 

In the straw-man's layered design, this creates a fundamental problem: by allowing the untrusted application to narrate its own actions in a separate log, the system lacks complete mediation for accountability from the outset. The continuity layer provides a trustworthy record of storage-level changes but is blind to their semantic meaning, making it impossible for auditors to attribute effects to their causes.

An untrusted application can manipulate its logs to mislead auditors by omitting, reordering, or misrepresenting operations. Such manipulations break three critical invariants: complete mediation (omitted entries make completeness unverifiable), single authoritative total order (reordered entries or manipulated timestamps create ordering ambiguity), and cryptographic binding between cause and effect (lies about operation scope make exclusivity and cross-object coherence unverifiable). Auditors thus cannot reliably prove facts about the history, as any result is best-effort and not cryptographically guaranteed to be correct.

\shortsection{Our Objective}
The analysis of the straw-man reveals two fundamental problems: a layered design makes accountability impossible, while rollbacks that rely on stale metadata break state continuity. Our objective is to solve both problems by moving the responsibility for managing and recording all state transitions into the continuity layer itself. We achieve this by making the continuity layer \emph{rollback-aware}, transforming rollback from an unmanaged application feature into a first-class, mediated, and auditable state transition. By unifying all state changes under this continuity layer, we satisfy the conditions for both state continuity (P1) and accountability (P3) by construction.

\subsection{Architecture}
\label{ssec:arch}
\begin{figure}[t]
\centering
\includegraphics[width=0.48\textwidth]{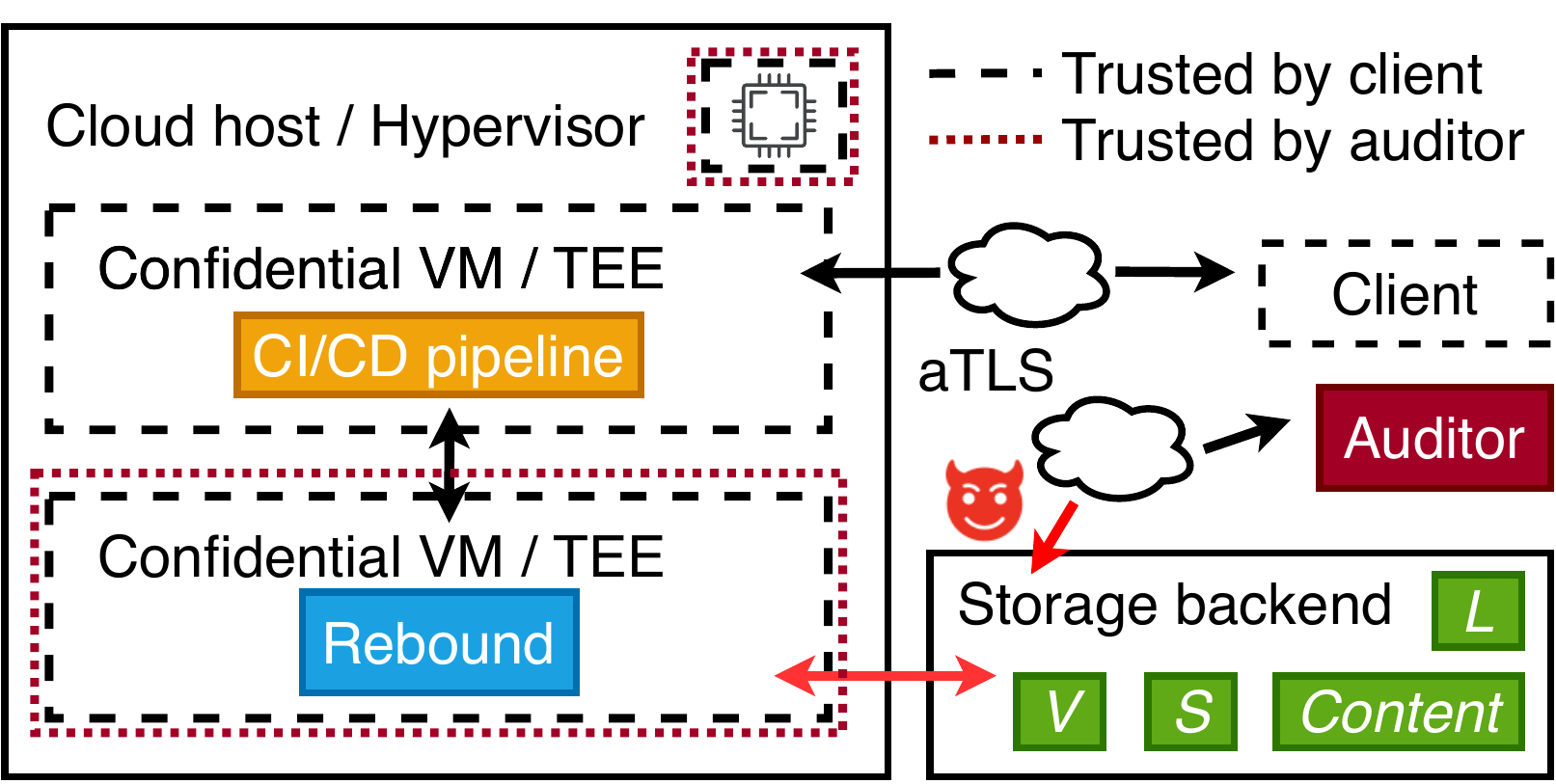}
\caption{\tool{} high-level system architecture.}
\label{fig:architecture}
\end{figure}

\tool{} mediates all state changes through a \emph{logically centralized} reference monitor that authorizes state changes and seals them in a single authoritative order. Clients and auditors attest and trust this reference monitor and only perform verifications against its sealed state. \autoref{fig:architecture} shows the high-level system architecture.

\shortsection{System Components}
Cloud applications using \tool{} are composed of six components:

\begin{description}
\item \textit{Application \& \tool{} Library:} Applications link to the library, which provides hooks to invoke \tool{} APIs by transferring control to the reference monitor.

\item \textit{Reference Monitor:} An isolated\footnote{Isolation can be achieved via TEEs or compartmentalization techniques.}, trusted service that executes the API calls, enforces policy decisions (i.e., eligible for rollback), and manages metadata (object digests, audit log) for clients and auditors. Invokes the hardware root of trust to seal the global digest.

\item \textit{Hardware Root of Trust:} Enforces hardware-based memory isolation and attestation of running application code. Also provides monotonic counters and sealing APIs via hardware-based cryptographic keys. If the root of trust supports a subset, \tool\ uses a TPM~\cite{shang2024ccxtrust}, HSM~\cite{thales2025monotonic}, or monotonic counter service~\cite{martin2021mcs,fernandez2024vmcaas}.

\item \textit{Version Catalog ($\mathcal{V}$):} Key-value index mapping (object ID, version ID) \(\rightarrow\) content digests, persisted on untrusted storage. KV pairs are verified by the reference monitor before use during state changes.

\item \textit{Snapshot Registry ($\mathcal{S}$):} Key-value index (tag \(\rightarrow\) version ids) persisted on untrusted storage. KV pairs are similarly verified by the reference monitor before use.

\item \textit{Audit Log ($\mathcal{L}$):} Append-only log of intents, decisions, \red{digests, and (optionally) authenticated identities}. Kept on untrusted storage but bound to the latest committed global digest. During audits, log contents are verified by auditors against the latest global digest sealed by the hardware root of trust.

\item \textit{Content Store:} Stores raw content on untrusted storage, indexed by content digest.
\end{description}

\shortsection{Workflow}
The typical workflow involves clients invoking \tool{}'s APIs through the application to manage state, with all operations mediated by the reference monitor. For a state update, the application computes content digests for the target objects (e.g., binaries, configuration files) and marshals an API request that maps each object to its digest. The reference monitor updates metadata and seals the new state. To create a snapshot (recovery point), clients specify a tag and the set of object IDs to include. For rollback, clients specify a snapshot tag or specific object versions to restore. For pruning, clients specify versions to remove from future use. Auditors can query the full, verifiable history of any object via a lineage reconstruction API.

In the CI/CD pipeline example, when a client merges a feature branch, the application builds the release bundle and deploys it, computes content digests, then invokes \tool{}'s state update API so the reference monitor seals the new state. If a bug is detected post-deployment, the client triggers a rollback job that redeploys the desired snapshot (e.g., \texttt{prod-stable-q2}), then calls \tool{}'s rollback API so the reference monitor seals the rolled-back state. 

%% file: 04.tex
\section{\tool{} Mechanisms}
\label{sec:protocols-impl}
This section presents the novel mechanisms behind the APIs in Section~\ref{sec:design}. At a high level, \tool{} retains classic forward-only sealing for rollback protection, but introduces a new abstraction layer that redefines \textit{what} is sealed to create a hardware-rooted, policy-enforcing state machine that enables secure rollback and pruning. \autoref{tab:notation-protocols} summarizes notation.

\subsection{Anchoring to an Authoritative Root}
\label{ssec:freshness-anchor}
As discussed in Section~\ref{sec:semantic-gap}, prior state continuity frameworks were not designed to support authorized rollbacks. They do not recognize them as legitimate operations nor provide a way to assert the freshness of historical metadata (or data). This presents a trustworthy oracle problem: how can historical metadata be securely verified?

\shortsection{Authentication Against the Latest Seal}
Addressing this requires that verifiers are able to request a single proof that asserts freshness and mutual consistency across all relevant sources. \tool{} enables such a proof by enforcing two invariants. First, all verifications of \textit{current} and \textit{historical} data and metadata are performed against a single global digest bound to the highest counter value. We refer to this digest (i.e., Merkle root) as the \textit{authoritative root} and denote it by $R$. Anchoring verifications to $R$ ensures that any data or metadata used in a decision is fresh. The reference monitor seals $R$ with the hardware monotonic counter, making it the sole source of truth for all verification decisions. Second, all protocols follow a two-step verification process. They first verify a digest $h$ against $R$ via an inclusion proof, then fetch the content, hash it, and verify the result matches $h$.

\begin{table}[t]
\centering
\small
\caption{Notation used in \tool{} mechanisms.}
\begin{tabular}{ll}
\toprule
\textbf{Symbol} & \textbf{Meaning} \\
\midrule
$O_i$ & Persistent object $i$ \\
$v_i^j$ & Version of object $O_i$ at counter $j$ \\
$h_i^j$ & Content digest of $v_i^j$ \\
$H_i$ & Current committed digest of $O_i$ \\
$\mathcal{V}$ & Object-Version Map (OVM) \\
$H_{\mathcal{V}}$ & Merkle root of OVM \\
$\mathcal{S}$ & Snapshot Registry \\
$H_{\mathcal{S}}$ & Merkle root of snapshot registry \\
$\mathcal{L}$ & Append-only audit log (intents/completions) \\
$H_{\mathcal{L}}$ & Merkle root of audit log $\mathcal{L}$ \\
$R$ & Authoritative root of the authenticated dictionary \\
$c$ & Current monotonic counter value \\
$\texttt{Seal}(x, c)$ & Sealing of $x$ bound to counter $c$ \\
\bottomrule
\end{tabular}
\label{tab:notation-protocols}
\end{table}

\shortsection{Security Guarantees}
This mechanism directly enforces our state continuity property (P1). Because all reads of data and metadata (e.g., snapshot listings) must be verified against the authoritative root $R$, any attempt to return stale content would result in a verification failure. This prevents both simple and second-order rollback attacks.

\subsection{Append-Only Rollback and Pruning}
\label{ssec:accumulating}
Anchoring all verifications to a fresh root $R$ ensures state continuity (P1) by preventing attacks that rely on stale metadata. However, this alone does not guarantee accountability (P3). The challenge is that modern systems require not just forward updates, but also rollback and pruning features, and the semantics of how these operations are implemented are critical.

\shortsection{Destructive Rollback and Pruning}
In current systems, both rollback and pruning are typically destructive. Rollback mechanisms either rewrite history (e.g., ZFS snapshots make post-rollback states inaccessible) or replace metadata in-place (e.g., software deployment pipelines overwrite image tags or re-tag manifests, losing the previous association). Pruning is similarly destructive. OCI registries delete tags and manifests. Kubernetes clusters garbage-collect old ReplicaSets. File systems simply remove historical snapshots.

While operationally convenient, these approaches create a fundamental \textit{accountability gap}. This gap exists even if we solve the layering problem (from the straw-man) and achieve complete mediation by moving all state management into the continuity layer (i.e., the trusted reference monitor). The issue is that destructive operations, by their very nature, erase the historical record. When the reference monitor executes a rollback or prune on behalf of a client, it leaves no durable evidence of the action. By erasing or altering history, the system makes it impossible for an auditor to later distinguish between legitimate, policy-driven changes and the effects of a compromised client. A compromised client could issue a rollback or prune request, and once the trusted monitor executes it destructively, there would be no cryptographic proof that the action ever occurred. Auditors would be unable to reconstruct what happened or attribute responsibility.

This is particularly dangerous for pruning. As noted in Section~\ref{sec:semantic-gap}, an attacker can mount a resurrection attack if effects of pruning decisions are not bound to the fresh anchor. However, even with a fresh anchor, if pruning is implemented as a simple deletion, it fails to create \textit{provable negative evidence}---a durable, authenticated record that a version was intentionally de-authorized. This breaks accountability (P3) in two ways. First, at rollback time, the reference monitor cannot cryptographically verify whether a version's absence from the eligible set is due to a legitimate pruning decision or a malicious omission by an attacker. Second, auditors reviewing the system logs after the fact cannot reconstruct what happened or attribute responsibility for the deletion.

\begin{figure}[t]
    \centering
    \includegraphics[width=0.4\textwidth]{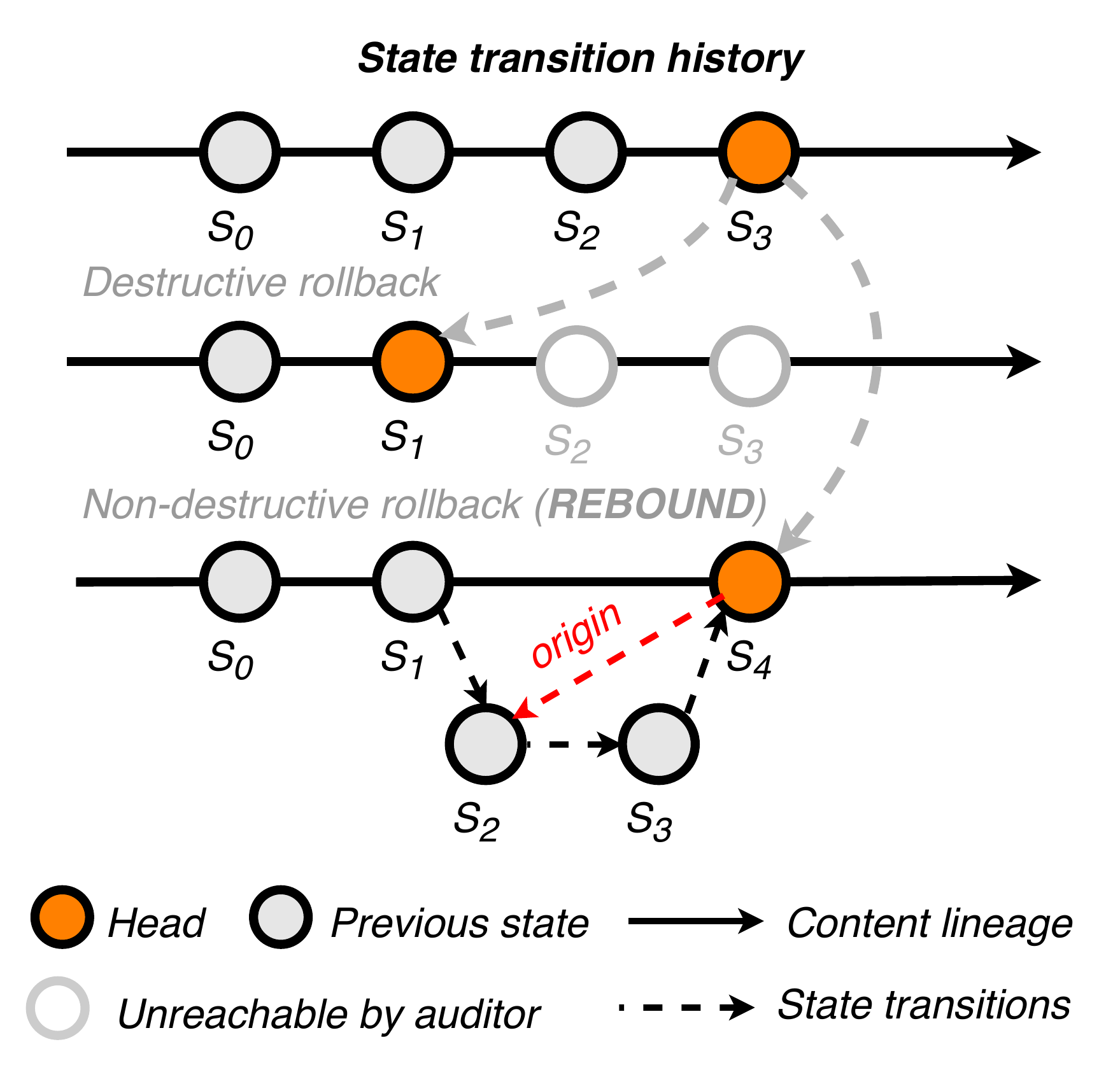}
    \caption{In \tool{}, rollback is represented and executed as a forward state transition that creates a new state and records lineage metadata (i.e., origin pointer) that can be used by auditors to reconstruct lineage.}
    \label{fig:rollback-protocol2}
\end{figure}

\shortsection{From Destructive Actions to Append-Only State Transitions}
To close this accountability gap, \tool{} transforms all operations into non-destructive, \textit{append-only state transitions}. Instead of deleting or overwriting, the system only ever adds new information. This ensures that the complete history of every action is preserved and verifiable. An overview is shown in Figure~\ref{fig:rollback-protocol2}.

To enable this, we build on a cryptographic primitive commonly used in certificate transparency (CT) systems: persistent authenticated dictionaries (PADs)~\cite{anagnostopoulos2001persistent}. A PAD is an append-only Merkle tree that indexes key-value pairs. It supports efficient \textit{membership verification queries}, allowing a verifier to cryptographically confirm whether a particular key-value pair exists in the dictionary (e.g., ``prove that version $v_i^j$ with digest $h_i^j$ is present''). The append-only property ensures that once a leaf is added, it remains in all subsequent versions of the tree, preventing retroactive modifications. 

What we need, however, goes beyond simple membership checks. We require a data structure that can accumulate the complete history of all state changes under the authoritative root $R$ and accumulate the \textit{policy decisions} that govern which historical states remain eligible for rollback. In other words, the structure must not only record what versions exist, but also encode authenticated metadata about their lifecycle status---i.e., active, rolled back, or pruned. This would allow the reference monitor to perform policy enforcement by consulting a single, fresh cryptographic anchor $R$. Note that we focus on enforcement; specifying policies is out of scope.

\shortsection{PAD Organization}
To achieve this, \tool{} organizes each of its three core state components---the Object-Version Map ($\mathcal{V}$), Snapshot Registry ($\mathcal{S}$), and Audit Log ($\mathcal{L}$)---as a separate PAD. Each state-changing operation appends new leaves to these structures and culminates in sealing a new root $R'$ with an incremented counter. For example, a state update appends entries to $\mathcal{V}$ mapping new version identifiers to content digests, optionally adds a snapshot entry to $\mathcal{S}$, and records intent and completion entries in $\mathcal{L}$. $\mathcal{V}$ also tracks the \textit{head} (i.e., the current live version) of each object. 

A rollback first verifies eligibility by checking inclusion in $\mathcal{V}$ and $\mathcal{S}$ under $R$ and ensuring no tombstone exists, then appends a new version entry (reusing a historical content digest) along with lineage metadata in $\mathcal{L}$. \red{A pruning operation verifies the pruning target under $R$ (either a snapshot tag in $\mathcal{S}$ for snapshot-based pruning, or a specific object version in $\mathcal{V}$ for selective pruning), then appends a \textit{de-authorization record} (a tombstone) marking it ineligible for future rollback.} This design echoes the use of tombstones in Dynamo/Cassandra~\cite{sivasubramanian2012amazon,hewitt2010cassandra} and other LSM-based stores~\cite{bailleu_speicher_nodate}, which employ similar markers to enable replica convergence and deferred garbage collection. By encoding rich metadata directly in the leaves (such as tombstones and lineage pointers), \tool{} transforms the PAD from a passive record into an active, policy-enforcing state machine.

\subsection{Co-Sealing History and Policy Decisions}
\label{ssec:co-sealing}
With the capability to accumulate history and policy decisions in append-only structures, the next question is the semantics of verification. Consider what the reference monitor must verify during a rollback to a tagged snapshot. As it reads from untrusted storage, it must query $\mathcal{S}$ to resolve the tag to a set of object versions, check each version digest's inclusion in $\mathcal{V}$, verify that no tombstones exist for any of these versions, and finally authenticate the actual content from the untrusted content store. Similarly, a state update requires verifying the current head of each object, while \red{pruning requires confirming the pruning target's existence (snapshot tag in $\mathcal{S}$ or version in $\mathcal{V}$) before appending the corresponding tombstone.} Each operation involves multiple inclusion proofs across different authenticated structures. The critical question is: how do we ensure these proofs are mutually consistent and bound to the freshest state?

\begin{figure}[t]
    \centering
    \includegraphics[width=0.45\textwidth]{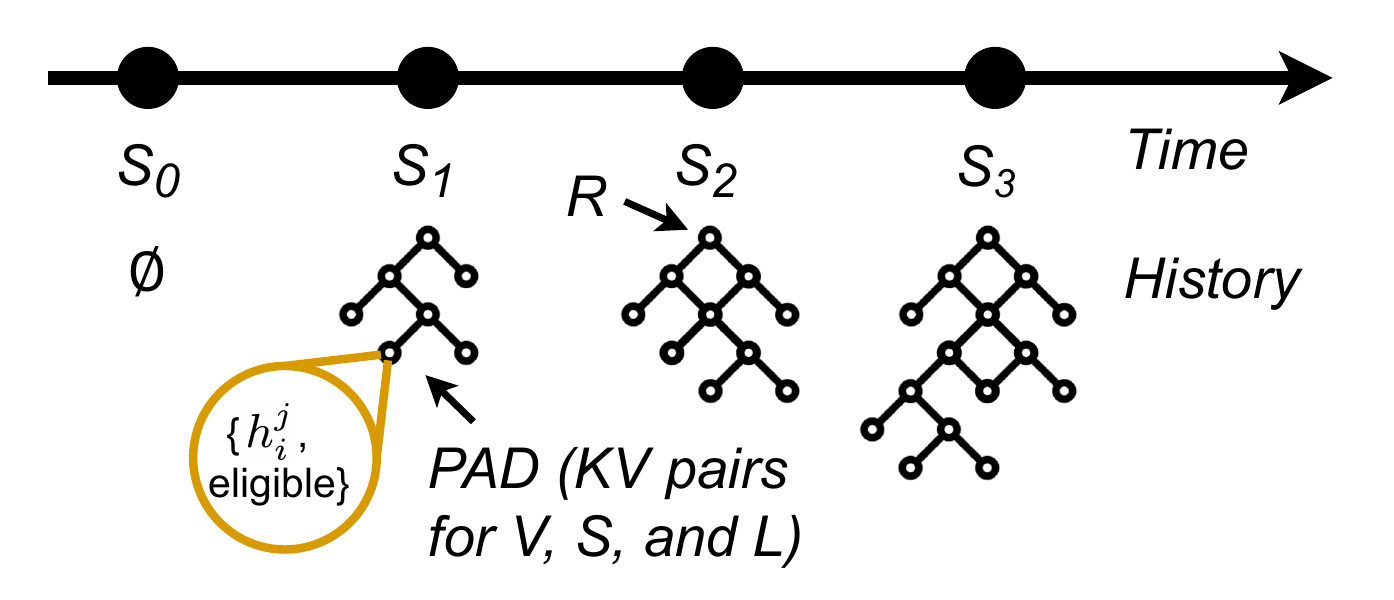}
    \caption{\tool{} uses a tailored authenticated dictionary to accumulate prior object versions and policy provenance under the authoritative root $R$.}
    \label{fig:pad-structure}
\end{figure}

Importantly, in \tool{}, the roots of these three PADs ($H_{\mathcal{V}}$, $H_{\mathcal{S}}$, $H_{\mathcal{L}}$) are aggregated into the single authoritative root $R$ (e.g., $R = \text{Hash}(H_{\mathcal{V}} | H_{\mathcal{S}} | H_{\mathcal{L}})$) and co-sealed with the hardware monotonic counter. An overview is shown in Figure~\ref{fig:pad-structure}. This seemingly simple architectural choice has significant implications: it transforms a collection of independent authenticated structures into a unified, policy-enforcing state machine where history and policy decisions are atomically bound under a single cryptographic commitment. This design stands in contrast to prior state continuity frameworks, which have not addressed how to bind policy decisions---such as de-authorization or lifecycle management---to a fresh anchor.

\shortsection{Unified Verification}
Co-sealing history and policy under a single root $R$ bound to a hardware monotonic counter enables unified verification: all queries are answered by one authoritative source with a single cryptographic commitment. The counter value $c$ serves as a universal binding point---given any sealed state at counter $c$, verifiers can cryptographically trace forward through version histories in $\mathcal{V}$, resolve snapshot membership via $\mathcal{S}$, and cross-check actions against audit records in $\mathcal{L}$, all anchored to $R$. As the system evolves and each new operation grows the PADs, a new $R$ is produced, ensuring that all historical and policy state remains cryptographically bound to the latest seal.

This approach enables three new accountability capabilities. First, \textit{eligibility verification}: the reference monitor performs multiple inclusion proofs (resolving snapshot tags in $\mathcal{S}$, retrieving versions from $\mathcal{V}$, checking for tombstones) all anchored to $R$, preventing resurrection attacks. Second, \textit{verifiable lineage reconstruction}: auditors can retrieve records from $\mathcal{L}$ and $\mathcal{V}$, verify all content digests, and confirm binding under $R$ (see Algorithm~\ref{alg:lineage}). Third, \textit{atomic policy enforcement}: because policy decisions are embedded in the authenticated structure rather than maintained externally, there is no window for decoupling freshness and compliance.

\shortsection{Security Guarantees}
Combined, our mechanisms directly enforce our observability and accountability property (P3) and satisfy design goals G1, G2, and G3. For P3, co-sealing $\mathcal{V}$, $\mathcal{S}$, and $\mathcal{L}$ under a single root $R$ bound to counter $c$ ensures that all state transitions are mediated by the reference monitor and recorded in a single authoritative order. The append-only nature provides complete mediation: all actions are captured as immutable records where tampering is detectable. The cryptographic binding between audit records in $\mathcal{L}$ (who/why) and version entries in $\mathcal{V}$ (what/when) ensures attribution is bound to exact effects, enabling auditors to verify completeness, exclusivity, and absence via succinct cryptographic proofs.

\shortsection{Functional and Performance Guarantees}
For design goals, the append-only structure enables authorized rollback (G1) by preserving historical versions under the fresh anchor, supports de-authorization via tombstones (G2) for compliance and removal of compromised versions, and provides efficiency (G3) through $O(\log n)$ query complexity in the PAD structure. Pruning also bounds storage costs by enabling deletion of content after tombstone records are established.

\begin{algorithm}[t]
\small
\caption{Lineage Reconstruction for $O_i$}
\label{alg:lineage}
\KwIn{Object $O_i$ and latest checkpoint $(\red{S_{\mathcal{L}}}, S, R, c)$, where \red{$S_{\mathcal{L}}$ is the number of leaves in $\mathcal{L}$ and} $S$ is the total number of leaves across all PADs}
\KwOut{Chronological sequence of version events with content lineage}
$\textit{events} \gets []$ \tcp{Accumulate lineage events}
$\textit{current} \gets \text{nil}$ \tcp{Track current head}
$\textit{hashMap} \gets \{\}$ \tcp{Map: content hash $\to$ first counter with that hash}
\For{each log entry $\ell$ at index $k \in [0, \red{S_{\mathcal{L}}})$}{
    Verify inclusion of $\ell$ at index $k$ under $R$ \tcp{All entries verified against current root}
    Parse $\ell$ to extract key-value pairs\;
    \If{$\ell$ contains head pointer for $O_i \to j$ \textbf{and} $j \neq \textit{current}$}{
        Extract from $\ell$: content digest $h_i^j$ for $(O_i, v_i^j)$ from $\mathcal{V}$\;
        Extract from $\ell$: metadata $\textit{txid}, \textit{desc}, \textit{origin}, \textit{ts}$ at counter $j$\;
        $\textit{contentOrigin} \gets \textit{hashMap}[h_i^j]$ \tcp{Find first occurrence of this content}
        \If{$\textit{contentOrigin}$ is nil}{
            $\textit{hashMap}[h_i^j] \gets j$ \tcp{First time seeing this content}
        }
        Append event: $(O_i, j, k, \textit{ts}, \textit{txid}, \textit{desc}, \textit{origin}, \textit{contentOrigin}, h_i^j)$\;
        $\textit{current} \gets j$\;
    }
}
\Return{$\textit{events}$}\;
\end{algorithm}

\subsection{Adversarially Robust Crash Recovery}
\label{ssec:atomicity}
\tool{} ensures secure recovery (P2) by enforcing two invariants. First, all state-changing operations follow transaction semantics. Unlike prior works which append single objects or blobs~\cite{bailleu_speicher_nodate}, \tool{}'s co-sealing of multiple authenticated structures ($\mathcal{V}$, $\mathcal{S}$, $\mathcal{L}$) and support for multi-object updates necessitate transactional atomicity across all affected leaves. \tool{} achieves this via write-ahead logging: every operation is logged as a pair of intent and completion records in the audit log $\mathcal{L}$. An operation is atomic if it either completes fully (producing a new authoritative root $R'$), or fails entirely (leaving the prior state $R$ as authoritative). When an operation begins, the reference monitor appends an intent record. The completion record is appended just before sealing so it is included under the new root $R'$. Because every operation culminates in sealing a single root, the newly sealed root is the sole commit point; until it is published, the prior state remains authoritative.

Second, if a crash occurs before completion, the system must recover to a consistent state. In the simplest case, the reference monitor scans $\mathcal{L}$ to identify the most recent transaction with both intent and completion records, which corresponds to the last fully-sealed root $R$. However, the untrusted storage can present ambiguous checkpoint scenarios: checkpoints ahead of the current counter value, multiple candidate checkpoints at different counter values, checkpoints behind the counter, or no valid checkpoint at all. \tool{}'s co-sealing mechanism enables administrators to make more informed, policy-aware recovery decisions than previously achievable. They can do this by inspecting and cross-checking the embedded metadata (version history, audit log, de-authorization records). We leave detailed investigation of the space of recovery policies to future work.

Once a recovery state is chosen, \tool{} applies the double-increment discipline from Ariadne~\cite{strackx2016ariadne}: the recovery protocol unconditionally performs a double-increment of the hardware counter, advancing from $c$ to $c+2$. This invalidates any speculative states staged for $c+1$ via prior crash attempts. Ariadne proved this mechanism is necessary to prevent attackers from using repeated crashes at select times to stage a partial, attacker-chosen state before it is sealed~\cite{strackx2016ariadne}; we refer to the Appendix for further details.

\shortsection{Security Guarantees}
These two invariants enforce secure recovery (P2). Transaction semantics ensure that only fully-sealed, completed transitions with matching intent-completion pairs are considered authoritative, while the double-increment discipline defeats attackers that attempt to stage partial state through repeated crashes.

%% file: 05.tex
\section{Security Analysis}
\label{sec:security-safety}

This section \red{informally} analyzes the security properties of \tool{} under the threat model defined in Section~\ref{sec:threat-model} and the goals defined in Section~\ref{sec:goals}. We consider two types of attackers: external (unauthorized) attackers who can cause system crashes or inject invalid data into the storage interface, and internal (authorized) attackers who can maliciously issue state changing commands via the reference monitor API. \red{Note that internal attackers can be compromised clients or administrators; while they can issue destructive commands (e.g., rollbacks, tombstones), they cannot hide or deny their actions because every operation is recorded in the audit log.}

\shortsection{Assumptions}
Our security analysis relies on standard cryptographic assumptions and correct hardware primitives: (1) collision-resistant cryptographic hash functions for Merkle tree construction and content addressing; (2) unforgeable sealing (MAC or digital signatures) provided by the hardware root of trust; (3) correct implementation of hardware monotonic counters that cannot be decremented or reset; and (4) correct memory isolation and attestation provided by the trusted execution environment. 

\subsection{Secure Recovery}

\noindent\textbf{\red{Property} 1 (Atomicity \& Secure Recovery).} \textit{At any point in time, the authoritative root reflects either the complete execution of an authorized operation or the unchanged prior root. No external or internal attacker can cause the system to commit a partially-applied operation.}\\

\noindent\textit{\red{Argument}.} Assume for contradiction that an attacker (external or internal) can cause the system to commit a root that represents a partially-applied operation, where some but not all components of a multi-object update or rollback have been applied. Consider an operation $\mathcal{O}$ that affects objects $O_1, \ldots, O_n$, transitioning the system from authoritative root $R_t$ sealed as $\texttt{Seal}(R_t, c_t)$ to target root $R_{t+1}$. \tool{} begins executing its atomic state update protocol to produce the new root $R_{t+1}$. For an attacker to succeed, the system must accept a sealed root $R'' \ne R_{t+1}$ at counter $c_t + 1$ that commits to only a strict subset of the intended changes.

\underline{\textit{External attacker.}} To achieve this, the attacker must either: (1) craft and inject $R''$ with a valid seal at $c_t+1$, which would require forging a seal (impossible under the unforgeability assumption of the hardware sealing primitive); or (2) crash the system during execution to publish a speculative partial root at counter $c_t+1$. However, the double-increment recovery discipline deterministically advances the counter from $c_t$ to $c_t+2$ after any crash, invalidating any speculative root at $c_t+1$ and forcing recovery to either the fully committed $R_{t+1}$ or the unchanged $R_t$.

\underline{\textit{Internal attacker.}} Internal actors must invoke the reference monitor's API. Even if an internal attacker intentionally attempts to produce a partial root (e.g., by triggering crashes at specific points), they cannot cause such a root to become authoritative because sealing is withheld until the operation completes fully. Either all components succeed and are sealed, or the operation fails and no seal is produced. This also ensures benign crash-consistency: after any crash, recovery validates the root against the current counter, ignoring uncommitted writes.

Thus, under standard cryptographic assumptions, attackers cannot cause the system to commit a partially-applied operation, contradicting our initial assumption.

\subsection{State Continuity}

\noindent\textbf{\red{Property} 2 (State Continuity).} \textit{At authoritative root $R_t$, no attacker (external or internal) can cause the system to accept a stale root $R_{t-k}$ (for any $k > 0$) as the current authoritative root.}\\

\noindent\textit{\red{Argument}.} Assume that an attacker can cause the system to accept a stale root $R_{t-k}$ as the authoritative root when the true authoritative root is $R_t$. Let $R_{t-k}$ be sealed as $\texttt{Seal}(R_{t-k}, c_{t-k})$ and $R_t$ be sealed as $\texttt{Seal}(R_t, c_t)$ where $c_t > c_{t-k}$ due to the monotonic counter property. We assume atomicity and crash-consistency as established in Property~1 and consider only steady-state acceptance of stale state.

\underline{\textit{External attacker.}} For the system to accept $R_{t-k}$ as authoritative, the attacker must replay an old seal $\texttt{Seal}(R_{t-k}, c_{t-k})$ (rejected by the monotonic counter property), forge a fresh seal $\texttt{Seal}(R_{t-k}, c_t)$ on stale content (infeasible under the unforgeability assumption), present stale data or metadata with valid inclusion proofs under $R_t$ (requiring hash collisions or preimages, which are infeasible under collision-resistance), or mount a second-order rollback attack (Section~\ref{sec:semantic-gap}) by presenting stale metadata with matching historical seals to mislead an administrator. The latter is defeated because the reference monitor verifies all metadata (snapshot listings, version catalogs) against the current authoritative root $R_t$ before doing rollbacks. Stale metadata cannot produce valid inclusion proofs under $R_t$.

\underline{\textit{Internal attacker.}} An internal attacker can invoke an authorized rollback via the reference monitor, but this creates a new authoritative root $R_{t+1}$ with counter $c_{t+1}$. The system never accepts the historical root $R_{t-k}$ as authoritative. If the attacker attempts to roll back to a de-authorized version, the eligibility check by the reference monitor (tombstone verification under $R_t$) blocks the operation, preventing resurrection attacks. Reintroducing a de-authorized version would require the tombstone entry to be absent from $\mathcal{V}$ under $R_t$, which would require forging $R_t$ or finding hash collisions (both infeasible). While an internal attacker may attempt to bypass the monitor and write directly to storage, such writes cannot produce a valid seal, and benign clients verify all state against the authoritative root $R_t$. Any bypass attempt will thus result in failed verifications, making unauthorized changes detectable and non-authoritative.

Thus, under standard cryptographic assumptions, attackers cannot cause the system to accept $R_{t-k}$ as authoritative, contradicting our initial assumption.

\subsection{Observability, Accountability, and Verifiable Lineage Reconstruction}

\noindent\textbf{\red{Property} 3 (Observability and Accountability).} \textit{Every state transition (update, snapshot creation, authorized rollback, or prune) generates verifiable audit records that cryptographically bind the operation's intent, scope, and outcome. No attacker (external or internal) can cause unobservable state changes or perform operations without generating verifiable audit records.}\\

\noindent\textit{\red{Argument}.} Assume that an attacker can cause a state transition from $R_t$ to $R_{t+1}$ without generating proper audit records, or can tamper with audit records to hide their actions. We rely on Property~1 for atomicity and crash-consistency and focus on steady-state verification.

\underline{\textit{External attacker.}} To succeed, an attacker must either: (1) forge a seal to publish a new authoritative root not accompanied by audit records, which would require forging $\texttt{Seal}(R_{t+1}, c_{t+1})$ without the reference monitor's cooperation (infeasible under the sealing unforgeability assumption); or (2) tamper with authenticated structures (the audit log $\mathcal{L}$, the OVM $\mathcal{V}$, or the snapshot registry $\mathcal{S}$) under $R_t$ so that records are absent, incomplete, or inconsistent, while reads still verify. However, any modification to these structures would change their Merkle roots ($H_{\mathcal{L}}$, $H_{\mathcal{V}}$, $H_{\mathcal{S}}$), which in turn changes the authoritative root $R_t$ itself (since $R_t = \text{Hash}(H_{\mathcal{V}} | H_{\mathcal{S}} | H_{\mathcal{L}})$). This would invalidate inclusion proofs under the original $R_t$ and require forging a new seal.

\underline{\textit{Internal attacker.}} All operations flow through the reference monitor, which atomically emits intent and completion records to $\mathcal{L}$, updates version entries in $\mathcal{V}$, and computes the new authoritative root $R_{t+1}$ before sealing. The cryptographic binding between audit records in $\mathcal{L}$ (capturing \emph{who} initiated the operation and \emph{why}) and version entries in $\mathcal{V}$ (capturing \emph{what} versions were affected and \emph{when}, indexed by counter value $c$) is enforced by co-sealing: all three structures are bound under a single root $R_t$ and sealed with the monotonic counter. This enables auditors to verify not only that an operation occurred, but also \emph{exactly which} objects were affected. For an internal attacker to suppress records or misrepresent the scope of an operation, they would need to tamper with authenticated state (requiring hash collisions or preimage attacks, which are infeasible) or forge a seal (also infeasible), reducing to the external case.

Therefore, under standard cryptographic assumptions, no attacker can cause unobservable state changes~\footnote{\red{Note that while all state changes are observable, internal attackers can still issue destructive commands within their authorization. Preventing abuse requires policy-level protections (e.g., 2-of-3 approval to execute tombstones, tombstone rate limiting, etc.), which we leave to future work.}} or operate without generating verifiable audit records, contradicting our initial assumption.

\medskip\noindent\textbf{\red{Property} 3.1 (Verifiable Lineage for Non-Linear Histories).} \textit{For any object version $(O_i, v_i^j)$ present at authoritative root $R$, its complete lineage is reconstructible with cryptographic proofs that bind each version's state, causal operation, and authorizing policy to a unique point in the system's history.}

\noindent\textit{\red{Argument}.} By Property~1, every operation is atomic and crash-consistent; by Property~3, every operation's intent and result are recorded in authenticated structures under the authoritative root $R$. The predecessor relation is recoverable from authenticated state: for ordinary forward updates, the predecessor of a newly created version $v_i^j$ is the prior head of $O_i$ at the immediately preceding counter value. For rollbacks, which create non-linear histories, the audit log records both intent (target snapshot or version set) and completion (newly created versions), and the version metadata includes an \texttt{origin} pointer linking $v_i^j$ to the historical version $v_i^k$ (where $k < j$) whose content is reused. This enables auditors to distinguish rollbacks from forward updates and reconstruct the full non-linear version graph.

Starting from $(O_i, v_i^j)$, an auditor reconstructs lineage by iteratively querying the audit log $\mathcal{L}$ for transaction records containing version transitions for $O_i$, extracting the predecessor pointer (or \texttt{origin} pointer for rollbacks) from each record's metadata, and verifying that all historical log entries are included under the current authoritative root $R$. Because all verifications are performed against $R$, any tampering with historical entries would change $H_{\mathcal{L}}$ and thus invalidate the seal on $R$. This ensures that the reconstructed lineage reflects an unbroken, append-only history where no entries were deleted, modified, or reordered.

To forge, hide, or reorder lineage entries, an attacker would need to tamper with the append-only log (changing $H_{\mathcal{L}}$ and thus $R$, invalidating the current seal and requiring hash collisions), forge inclusion proofs for non-existent entries under $R$ (requiring preimages), or selectively omit historical entries (detectable because the current $R$ commits to all entries via $H_{\mathcal{L}}$). All such attacks require breaking collision-resistance of the hash function or forging seals.

%% file: 06.tex
\section{Performance Evaluation}
\label{sec:evaluation}

We evaluate \tool{} across a set of macro-benchmarks and micro-benchmarks. Like prior works~\cite{niu2022narrator,angel2023nimble}, we measure \textit{operation latency} and \textit{storage consumption} overheads. Our evaluation focuses on two questions:
\begin{enumerate}
    \item \textit{What end-to-end overheads does \tool{} add to real-world CI workflows?}
    \item \textit{What are the fundamental latency and storage overheads of \tool{}, and how do they scale?}
\end{enumerate}

\shortsection{Implementation}
We implement \tool{} in 4K lines of Go. The core library can be linked directly into applications or deployed as a standalone service with a REST API frontend. The implementation provides four main APIs as described in Section~\ref{ssec:arch}: \texttt{state\_update}, \texttt{take\_snapshot}, \texttt{rollback}, and \texttt{prune}. Both rollback and prune support two modes: snapshot-wide operations and selective operations on specific object sets.

\begin{figure}[t]
    \centering
    \includegraphics[width=0.4\textwidth]{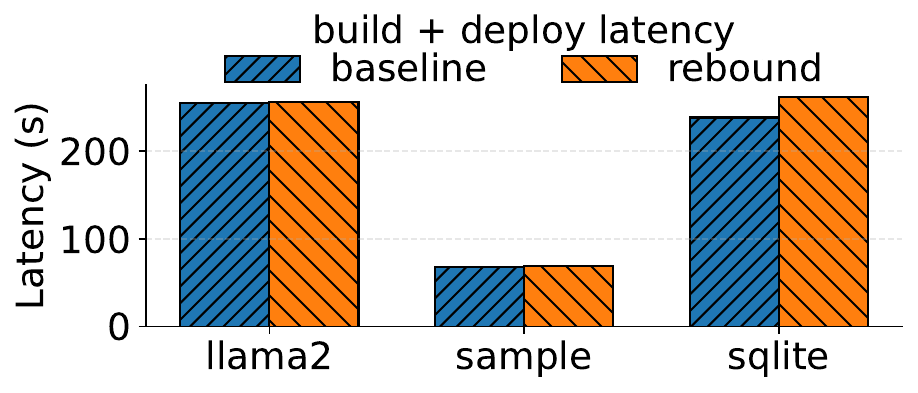}
    \includegraphics[width=0.4\textwidth]{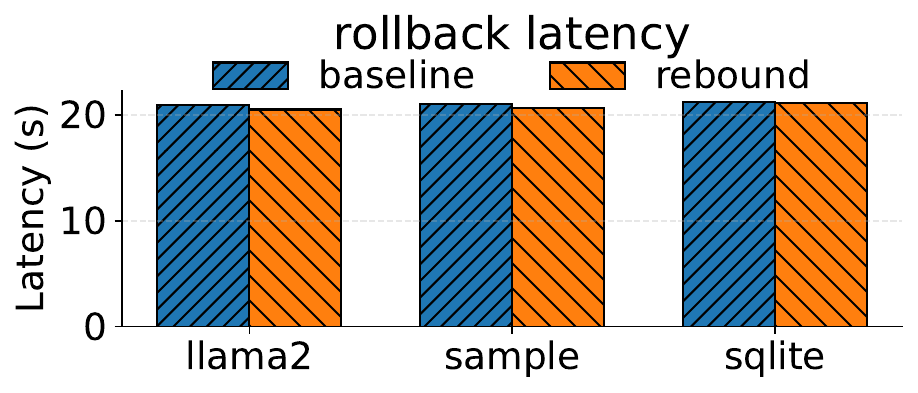}
    \caption{End-to-end latency for build/deploy and rollback jobs. \tool{} adds only seconds of overhead to complex builds that take minutes (or hours), making cryptographically verified deployments and rollbacks practical.}
    \label{fig:e2e-build-deploy}
\end{figure}

We use Google's Tessera library~\cite{tessera-proj,eijdenberg2015verifiable} to implement the underlying PADs, aggregating the version catalog ($\mathcal{V}$), snapshot registry ($\mathcal{S}$), and audit log ($\mathcal{L}$) into a single digest that is sealed by the hardware. Clients (operators, administrators, or other applications) connect over attested TLS to invoke state-changing APIs (\texttt{state\_update}, \texttt{take\_snapshot}, \texttt{rollback}, \texttt{prune}), while auditors use read-only APIs (\texttt{list\_snapshots}, \texttt{reconstruct\_lineage}) to query historical facts against the sealed authoritative root.

\shortsection{Experimental Setup}
We run all experiments on AWS EC2 m6a.2xlarge instances with 8 vCPUs, 32\,GB RAM, and Ubuntu 24.04 LTS with AMD SEV-SNP enabled. All application and \tool{} code thus runs inside confidential VMs. \tool{} uses a local ext4 file system as its storage backend; any backend could be used (e.g., AWS S3).

\subsection{Macro-benchmarks}
The macro-benchmarks invoke real GitLab CI pipelines for the popular open-source projects \texttt{llama2.c} (an AI inference engine, 3K lines of C/Python) and \texttt{sqlite} (a SQL database, 300K lines of C), and a minimal Go application (10 lines of Go) to isolate network and orchestration overheads from \tool{}'s cryptographic operations. We stood up a multi-container application (containing a gitlab-ce, gitlab-runner, Kubernetes controller, and \tool{} service), execute end-to-end CI pipelines (build, deploy, etc.), and measure end-to-end overheads (which includes network latency). For reproducibility, we use GitLab CI with locally-hosted runners and port representative build, deploy, and test phases from the projects' existing workflows. 

Integrating \tool{} is straightforward: new jobs are added to the YML files that invoke \tool{}'s APIs before or after existing build/deploy jobs. In our workloads, this amounted to approximately 70 lines of YAML per job for each of the five operations (state update, take snapshot, rollback, prune, and audit lineage reconstruction). Each macro-benchmark is run 10 times, and we report the average end-to-end latency for each job.

Each workload deploys 5--25 versioned objects per update. This range reflects the structure of modern software releases: following microservices best practices~\cite{dragoni_microservices_2016,chen_microservices_2018}, each service is independently deployable with (1) one or more container images, (2) Kubernetes manifests (Deployment, Service, ConfigMap, Secret, optionally Ingress), (3) configuration files, and (4) supply chain metadata required by frameworks like SLSA and in-toto~\cite{torres2019toto} (SBOMs, provenance attestations, signatures). A single-service deployment thus comprises \(\sim\)5--10 objects, while complex releases with multiple interdependent services or additional artifacts (e.g., database migrations, feature flags) range up to 25 objects. Atomic versioning of these interdependent components is critical for rollback correctness---reverting a container image without its matching configuration or SBOM would violate deployment integrity and compliance requirements.

\begin{figure}[t]
    \centering
    \includegraphics[width=0.5\textwidth]{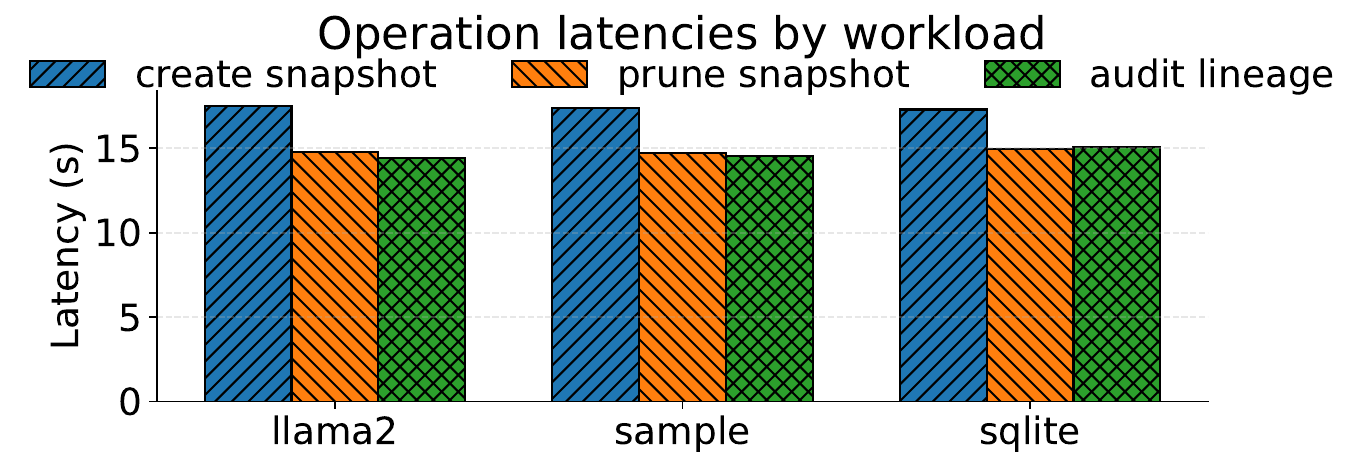}
    \caption{End-to-end latency for maintenance jobs. Snapshot creation, pruning, and audit operations all complete in <17 seconds. They are typically offline tasks and thus enable software deployment maintenance without meaningfully impacting developer workflows.}
    \label{fig:e2e-maintenance}
\end{figure}

\shortsection{Results: Build and Deploy Latency}
\autoref{fig:e2e-build-deploy} (top) shows end-to-end latency for the build and deploy pipeline, comparing \tool{} against the baseline. The x-axis represents the three workloads, while the y-axis shows latency in seconds. For \tool{}, the measured time includes building the application, deploying to Kubernetes, \textit{and} calling the \texttt{state\_update} API, which internally computes inclusion proofs and updates the version catalog, snapshot registry, and audit log data structures. 

\tool{} adds $<1$ second of overhead for \texttt{llama2.c} and \texttt{sample}, and approximately 10 seconds for \texttt{sqlite} (220 seconds baseline vs. 230 seconds with \tool{}, or roughly <5\% overhead). The key insight is that \tool{}'s overhead is modest even for our test workloads that have build/deploy latencies in minutes. For production builds of large repositories (e.g., Chromium with 36+ million lines, Android with 250+ million lines) that routinely take hours, \tool{}'s single-digit-second overhead for cryptographic state tracking and hardware-rooted integrity provides strong security guarantees without materially impacting developer productivity or CI pipeline throughput.

\shortsection{Results: Rollback Latency}
\autoref{fig:e2e-build-deploy} (bottom) shows end-to-end rollback latency across the same three workloads. The baseline measures the time to select a prior GitLab release and redeploy to Kubernetes. For \tool{}, the measured time includes calling the \texttt{rollback} API (which verifies snapshot eligibility, reconstructs the target state via inclusion proofs, appends new version entries with lineage metadata, and seals the updated root) plus the Kubernetes redeployment. 

\tool{} adds under 1 second of overhead across all workloads, with total rollback time of approximately 20 seconds. This demonstrates that \tool{} makes cryptographically verified, hardware-attested rollback---including full lineage reconstruction and tamper-proof audit trails---cost essentially the same as manual rollback procedures that offer no integrity guarantees. Organizations can adopt strong security controls for incident response without sacrificing the speed required during outages or security incidents.

\begin{figure}[t]
    \centering
    \includegraphics[width=0.45\textwidth]{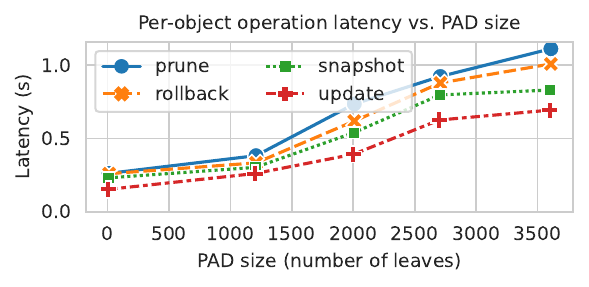}
    \caption{Operation latencies vs. number of leaves in the PAD. All operations exhibit sublinear scaling, completing in \red{0.5--1} seconds per object even in a PAD with 2,700 leaves. This shows that \tool{} mechanisms are practical in the context of the software deployment pipeline: with 25 objects grouped into updates, this corresponds to approximately 100 software releases.}
    \label{fig:micro-vs-leaves}
\end{figure}

\shortsection{Results: Maintenance Operations}
\autoref{fig:e2e-maintenance} shows the latency of maintenance operations across workloads. All three operations complete in under 17 seconds across all workloads, with snapshot creation being the fastest (\(\approx\)15 seconds) and audit operations taking slightly longer (\(\approx\)16 seconds) due to lineage traversal. Critically, these operations are often performed offline or as scheduled background tasks, rather then on the critical deployment path, and thus have more relaxed latency expectations. 

The key insight is that \tool{} enables continuous compliance and forensic capabilities without interfering with time-sensitive CI workflows. This includes complete audit log verification and storage reclamation via pruning. Even with relaxed expectations, these operations remain fast enough for frequent execution (e.g., hourly snapshots, daily audits).

\begin{figure}[!t]
    \centering
    \includegraphics[width=0.4\textwidth]{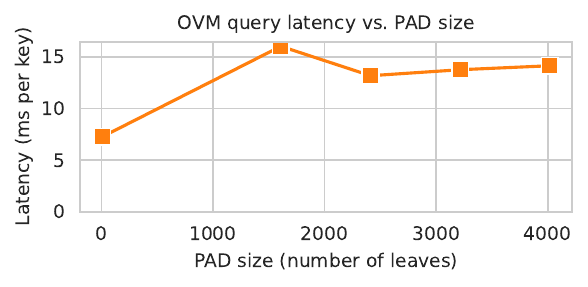}
    \includegraphics[width=0.4\textwidth]{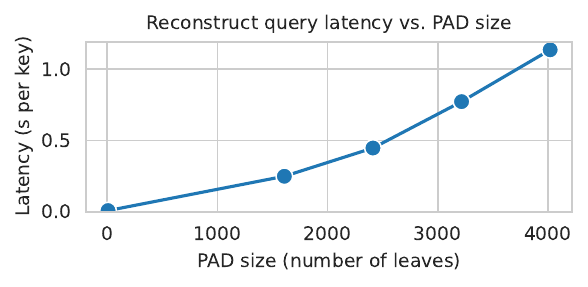}
    \caption{Query latency vs. history length. OVM queries scale logarithmically (<13ms at 2,700 leaves), while lineage reconstruction shows $O(k \log n)$ growth but remains practical for offline audit (<600ms).}
    \label{fig:query-vs-history}
\end{figure}
\begin{figure}[t]
    \centering
    \includegraphics[width=0.45\textwidth]{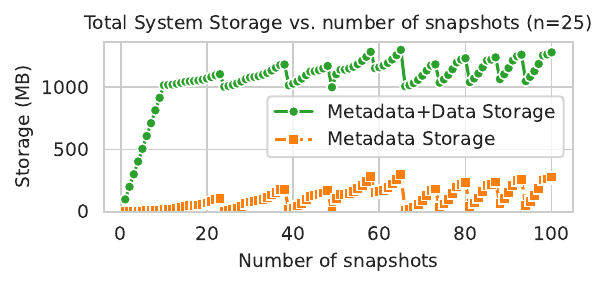}
    \caption{Bounded history accumulation and storage costs with pruning. Total storage can be capped by pruning raw data while retaining full cryptographic accountability over metadata.}
    \label{fig:storage-vs-history}
\end{figure}

\subsection{Micro-benchmarks}
The micro-benchmarks stress individual API operations. These experiments isolate the fundamental overheads of \tool{}'s cryptographic core, without network or CI orchestration costs, to reveal its scaling properties. Each experiment was run 10 times, and we report the average results. Note that all experiments run on AMD SEV-SNP confidential VMs, which impose a baseline performance penalty of approximately 50\% compared to non-confidential environments due to memory encryption overheads~\cite{yan2023performance}.

Latency results are plotted against the \textit{number of leaves in the PAD}. Note that the PAD is based on a Merkle tree. The total number of leaves is determined approximately by the product of the number of updates and the number of objects per update (see Table~\ref{tab:analytical-costs}). This metric unifies two key dimensions of workload scale: when we fix the number of objects per update (e.g., 25) and vary the number of updates, the x-axis directly measures \textit{history length}. Conversely, when we fix the history length at a single update and vary the number of objects, the x-axis measures \textit{state size}. This representation allows us to reason about performance as a function of the total PAD size, regardless of whether that size results from many small updates or fewer large ones. For all latency measurements, we thus measure per-object costs (i.e., $N=1$), which can be extrapolated to larger object sets by multiplication. The Appendix provides a detailed breakdown of the analytical costs per operation, including leaf counts, seal operations, and \red{number of proofs}.

\shortsection{Operation Latency vs. PAD Size}
\autoref{fig:micro-vs-leaves} shows per-object operation latency as a function of PAD size for the four core operations. All operations exhibit sublinear scaling behavior. To contextualize: after 100 releases of 25 objects each (approximately 2,700 leaves per Table~\ref{tab:analytical-costs}), all operations complete in \red{0.5--1} seconds per object. \red{Note that state updates complete within 400ms per object at a history length of zero; for reference, prior works such as Nimble (which do not support rollback) report state update latencies of 40ms per object~\cite{angel2023nimble}. We refer to Table~\ref{tab:rebound_vs_nimble} for further details on prior works and Figure~\ref{fig:micro-concurrency} for details on concurrency.}

We note that the absolute latencies are largely constrained by limitations in the Tessera library. We use this library because of its prevalence (i.e., also used internally by Sigstore~\cite{newman2022sigstore}). However, we found its performance limited for our multi-object transactional use case, whereas prior works primarily dealt with non-transactional, single-object updates. Namely, hardcoded library parameters restrict the frequency of sealing. Fortunately, recent works on high-performance Merkle tree implementations have demonstrated that updates over large object sets can complete in hundreds of microseconds~\cite{burke2025efficient}, suggesting substantial room for optimization. We leave this to future work.

Beyond implementation improvements, there are design trade-offs in how objects are packed into transactions. Organizations could aggregate multiple object digests into fewer transactions to limit history growth, or commit PAD updates asynchronously to decouple operation latency from seal frequency. Log rotation techniques (common in CT systems) could periodically start fresh PADs, though this introduces trade-offs in verification complexity. The key result is that transaction latency remains practical even as history accumulates, and there are multiple paths for further optimization depending on deployment requirements.

\shortsection{Query Latency vs. PAD Size}
\autoref{fig:query-vs-history} shows the latency for read-only queries as history grows. OVM queries are the fundamental mechanism through which the APIs are executed: they provide the inclusion proofs to verify anything in the PADs, which are used internally to verify entries in a snapshot, check for tombstones, etc. We observe that OVM queries exhibit logarithmic growth, at <13 milliseconds even at 2,700 leaves, consistent with Merkle tree inclusion proofs that scale with tree height rather than size.

In contrast, full lineage reconstruction queries show super-linear growth, reaching approximately 600 milliseconds at 2,700 leaves. This is expected: \red{reconstructing the complete modification history for an object requires scanning all $S_{\mathcal{L}}$ entries in the audit log $\mathcal{L}$ to find those belonging to the object, verifying an inclusion proof in $\mathcal{L}$ under the authoritative root for each. Since each proof costs $O(\log S_{\mathcal{L}})$, the total complexity is $O(S_{\mathcal{L}} \log S_{\mathcal{L}})$, which is reflected in the observed curve.} Despite this growth, the absolute latencies are practical for audits, which are typically performed offline.

The key insight is that \tool{} achieves the expected asymptotic behavior: cryptographic verification operations scale logarithmically. As discussed, the absolute latencies are constrained by implementation factors discussed above (CVM overhead, Tessera implementation limits).

\shortsection{Storage vs. History Size}
\tool{}'s storage costs are driven by two components: the append-only metadata log and the versioned object data itself. \autoref{fig:storage-vs-history} illustrates this trade-off. In this experiment, pruning is configured to keep the 10 most recent snapshots, with each snapshot comprising approximately 100MB across all 25 versioned objects. This represents a conservative estimate: real-world software deployments typically include at least one large container image artifact (often multi-gigabyte in size), along with Kubernetes manifests, configuration files, and supply chain metadata. The plot shows that while metadata storage grows unboundedly, the total system storage can be effectively capped. 

We observe that the raw data storage stops growing once the 10-snapshot limit is reached. From that point, the total storage is driven entirely by the much smaller metadata footprint. This demonstrates a critical capability: organizations can achieve full cryptographic accountability---maintaining an immutable audit trail that proves what data existed at every point in history---while bounding storage costs by aggressively pruning the actual raw data artifacts. This separation of metadata (for accountability) and data (for operational needs) enables cost-effective compliance: organizations can retain verifiable proof of historical deployments without retaining the artifacts themselves indefinitely.

\begin{tcolorbox}[colframe=white,colback=gray!10]
\textbf{Evaluation Summary:} \tool{}'s security mechanisms incur modest overhead: adding only <5\% latency to complex CI workloads and enabling sub-second audit queries. These guarantees come at a cost that is negligible for production deployments, making cryptographically verifiable version control practical for real-world software deployments.
\end{tcolorbox}

%% file: 07.tex
\section{Related Work}
\label{sec:related}

\tool{} draws on ideas from state continuity frameworks, verifiable audit logging, and version control systems.

\shortsection{State Continuity and Authenticated Storage}
As discussed in Section~\ref{sec:background}, prior frameworks like Memoir~\cite{parno2011memoir}, Ariadne~\cite{strackx2016ariadne}, ROTE~\cite{matetic2017rote}, and Nimble~\cite{angel2023nimble} guarantee freshness: no stale data is ever recognized as current. These systems build on authenticated storage primitives (e.g., hash chains)~\cite{merkle1989certified,sinha2018veritasdb,arasu2021fastver,tamassia2003authenticated,erway2015dynamic,naor2000certificate,miller2014authenticated} and TEEs~\cite{mckeen_innovative_2013,aws_sev_snp,baumann_shielding_2015,sabt_trusted_2015,ngabonziza_trustzone_2016,van2019sok,Costan2016IntelSE} that provide sealing, attestation, and monotonic counters. However, they treat all rollbacks as adversarial and provide no mechanism for authorized rollbacks. \tool{} extends these seminal works by making rollback a first-class, policy-governed operation while preserving the core freshness guarantee.

\shortsection{Verifiable Causality and Audit Logging}
Tamper-evident logs~\cite{crosby2011authenticated,parkinson2016auditing,paccagnella2020custos,ahmad2022hardlog} and provenance systems~\cite{inam2023sok,hassan2020omegalog,pasquier2018runtime,jamil2018secure,li2021threat} record event causality, typically focusing on system calls, process execution, or file accesses. These works primarily focus on data representation, auditing efficiency, and preventing log tampering in different threat models (based on TEEs or secure coprocessors). Systems like Bonsai~\cite{gehani2007bonsai} provide cryptographic lineage authentication over audit logs. They similarly rely on append-only data structures to enable efficient verification of event histories. \tool{} enables a fundamentally different class of queries: rather than tracking \textit{how} programs execute, it provides cryptographic guarantees about \textit{persistent state evolution}, including queries about data versions, rollback events, and pruning decisions. 

Critically, \tool{} supports queries over \textit{non-linear histories}---where state can move backward via rollback or be selectively pruned---which these prior works do not provide cryptographic guarantees for. Provenance systems focus on tracking execution causality (how programs run), not persistent state transitions (how data evolves). Rollback of persistent data is outside their threat model entirely.

\shortsection{Version Control and Recovery}
Git~\cite{torvalds2010git} provides flexible versioning for source code, and in spirit, \tool{} aims to provide similar usability for persistent state management. However, Git has no rollback protection: without hardware-rooted sealing and monotonic counters, it cannot prevent rollback attacks when state is stored on untrusted media, as adversaries can replay stale repository state and present it as current. Moreover, Git's design permits destructive operations (rebase, filter-branch, reflog pruning) that erase history and violate accountability requirements. \tool{} preserves Git-like flexibility (selective rollback, snapshot tagging, lineage queries) but enforces an append-only discipline with authenticated dictionaries and sealed roots. History cannot be rewritten; rollbacks and pruning are explicit, policy-checked operations that leave cryptographic evidence, making \tool{} suitable for high-assurance environments where accountability is mandatory. 

\red{One might envision a state-continuity-protected Git by simply disabling history-rewriting commands (e.g., \texttt{rebase}, \texttt{filter-branch}, \texttt{force-push}) via server-side hooks. However, disabling these operations is insufficient: the issue is not which commands are permitted but that Git's rollback model is semantically destructive by design. A \texttt{git reset} or \texttt{git rebase} rewrites history in-place rather than producing a new forward state transition. Even with destructive commands disabled, Git has no mechanism to represent a rollback as an explicit, append-only, auditable event---it cannot encode the distinction between ``this state was intentionally rolled back to version $v$ under policy'' and ``this state was corrupted or tampered with.'' \tool{} resolves this by treating rollback as a first-class forward state transition that appends to history rather than erasing it.}

Moreover, database point-in-time recovery systems~\cite{verhofstad1978recovery} and filesystem snapshots lack hardware-rooted anchors and cannot provide cryptographic guarantees about rollback authorization or accountability---they fall under the straw-man design described in Section~\ref{sec:semantic-gap}, where rollback utilities operate without a fresh authentication anchor linking current state to historical policy decisions. 

\red{\shortsection{Certificate Transparency} Certificate Transparency (CT)~\cite{laurie_rfc_2013} uses persistent authenticated dictionaries (PADs)~\cite{anagnostopoulos2001persistent}---the same append-only Merkle tree primitive at the core of \tool{}---to provide tamper-evident public logs of TLS certificate issuance. Broadly speaking, CT's security relies on a federated ecosystem of independent log operators, monitors, and auditors that gossip about Signed Tree Heads to detect equivocation. \tool{} occupies a fundamentally different point in the design space: rather than distributing trust across a public PKI federation, it uses a hardware monotonic counter as a closed-loop freshness anchor for a single cloud application. This eliminates the need for gossip protocols, multiple log operators, or out-of-band STH distribution; the counter provides the equivalent of equivocation detection via a single trusted hardware primitive. CT is also designed exclusively for forward-only certificate issuance; rollback has no analogue in CT's use case. \tool{} extends the PAD primitive to support non-linear histories where rollback is a first-class, policy-governed operation rather than an anomaly.}

%% file: 08.tex
\section{Conclusion}
\label{sec:conclusion}
We presented \tool{}, a framework that reconciles the security guarantees of state continuity with the operational need for authorized rollbacks. \tool{} establishes a new security primitive: hardware-rooted version control for persistent state. This enables powerful new security paradigms, such as allowing systems to move beyond point-in-time remote attestation to a model of continuous attestation, where TEEs can cryptographically prove their entire history of state transitions. This capability provides a foundation for provably correct distributed workflows and non-repudiable accountability for autonomous systems. As cloud applications grow in complexity and autonomy, we believe such a paradigm will become a cornerstone of trustworthy computing.

%% file: acks.tex
\section*{\red{Acknowledgments}}
\red{We thank the anonymous reviewers and shepherd for their insightful feedback. This work was supported in part by PRISM, one of seven centers in JUMP 2.0, a Semiconductor Research Corporation (SRC) program sponsored by DARPA.}

%% file: appx-1.tex
\section*{Appendix}

\section*{Crash Recovery \& Double-Increment Discipline}
The double-increment discipline prevents a specific class of hardware-based crash attacks. Consider an attacker who can trigger crashes (e.g., via power faults or injected exceptions) at precise moments during a rollback or pruning operation. Without the double-increment, the attacker could force the reference monitor to stage a partial state at counter $c+1$---for example, updating the heads in $\mathcal{V}$ to point to a historical version without first verifying that no tombstone exists for it, or appending a rollback completion record without actually performing the eligibility check. By repeatedly crashing and restarting, the attacker attempts to get this partial state sealed. The double-increment discipline defeats this by invalidating any state associated with $c+1$ during recovery, ensuring that only fully-sealed, completed transitions with matching intent-completion pairs are ever considered authoritative.

\section*{\red{Considerations for Low-Latency, High-Throughput Environments}}

\red{\tool{}'s current implementation builds on Tessera, a tile-based transparency log framework well-suited to our primary case study of CI/CD pipelines, where state updates are relatively infrequent (on the order of deployments). However, applications with more frequent state updates---such as high-throughput databases, real-time telemetry pipelines, or distributed key-value stores---will require a highly performant continuity layer capable of sustaining significantly higher update rates. Supporting such regimes will likely require replicating or sharding the reference monitor and its underlying authenticated structures ($\mathcal{V}$, $\mathcal{S}$, $\mathcal{L}$) across multiple nodes, with mechanisms to synchronize state updates and maintain a consistent authoritative root across replicas. Beyond conventional distributed systems challenges of replication and consensus, these environments introduce new security problems specific to the continuity layer: ensuring that no replica can diverge, equivocate, or accept stale state without detection, and that monotonic counter semantics are preserved in the presence of concurrent writers and partial failures. These challenges go beyond what is addressed by standard distributed systems techniques, and we leave them to future work.}

\section*{\tool{} Protocols}
\label{sec:protocols-detailed}

\begin{figure}[!h]
\centering
\small
\begin{mdframed}[linewidth=1pt,linecolor=gray!30,backgroundcolor=gray!5]
    \textbf{State Update Protocol.} \textit{Input:} current authoritative state $(R,c)$, set of modified objects $\Delta = \{O_i\}$ with proposed new bytes; optional snapshot designation with tag $t$ and object set $S_t$. \textit{Output:} new authoritative state $(R', c', \sigma')$ and authenticated log records. The reference monitor (RM) executes:
\begin{enumerate}[leftmargin=1.1em, itemsep=2pt]
    \item \textbf{Intent Declaration.} Generate a fresh transaction identifier $txid$; append intent record to audit log $\mathcal{L}$ containing prior root $R$, affected object identifiers, and (optionally) snapshot tag $t$ with the counter-indexed versions it would bind.
    \item \textbf{Policy \& Integrity Validation.} Verify proposer authorization, required approvals, time constraints, object dependency/consistency rules, and integrity of proposed bytes (hash, format, signature checks). Abort on any failure.
    \item \textbf{Per-Object Digest Preparation.} For each $O_i \in \Delta$, compute new content digest $h_i^{new}$. The version identifier is the next counter value $c' = c+1$.
    \item \textbf{Version Catalog Extension.} Append new entries to $\mathcal{V}$ mapping each $(O_i, c')$ to $h_i^{new}$ and update the head pointer for each $O_i$ to $c'$. Record the prior head value as the \texttt{origin} pointer in the version metadata, establishing the predecessor relation for lineage reconstruction.
    \item \textbf{Optional Snapshot Registration.} If requested, append entry to $\mathcal{S}$ binding tag $t$ to the counter-indexed versions (object-counter pairs).
    \item \textbf{Completion Logging.} Append completion record to $\mathcal{L}$ linking $txid$, prior root $R$, and the affected objects with their counter-indexed versions.
    \item \textbf{Sealing.} Compute roots $H_{\mathcal{V}}$, $H_{\mathcal{S}}$, $H_{\mathcal{L}}$ of the three PADs, aggregate them into new authoritative root $R'$, and obtain seal $\sigma' = \texttt{Seal}(R', c')$ where $c' = c+1$.
    \item \textbf{Publication.} Durably persist $(R', c', \sigma')$, then advance the hardware monotonic counter from $c$ to $c'$ to finalize $(R,c) \leftarrow (R', c')$, making $R'$ authoritative.
\end{enumerate}
\end{mdframed}
\caption{State update protocol: appends to $\mathcal{V}$, optionally $\mathcal{S}$, and $\mathcal{L}$, then co-seals under $R'$.}
\label{fig:state-update-protocol}
\end{figure}

\begin{figure}[!h]
\centering
\small
\begin{mdframed}[linewidth=1pt,linecolor=gray!30,backgroundcolor=gray!5]
	\textbf{Rollback Protocol.} \textit{Input:} current authoritative state $(R,c)$; rollback request specifying either snapshot id $s$ or explicit target set $T = \{(O_i, k_i)\}$ where $k_i$ denotes historical counter values, plus justification. \textit{Output:} new authoritative state $(R', c', \sigma')$ and authenticated log records. Note: systems typically operate in one mode (snapshot-based or selective) determined by policy; supporting both requires additional consistency checks.
\begin{enumerate}[leftmargin=1.1em, itemsep=2pt]
    \item \textbf{Intent Declaration.} Generate a fresh transaction identifier $txid$; append intent record to $\mathcal{L}$ with prior root $R$, mode (snapshot-based or selective), snapshot id or explicit target set, current heads of affected objects, actor, and justification.
  \item \textbf{Target Resolution.} If snapshot-based, verify snapshot binding in $\mathcal{S}$ under $R$ and resolve $s$ to a target set $T = \{(O_i, k_i)\}$; if selective, use the caller-provided $T$.
  \item \textbf{Eligibility Validation.} Check approvals and time windows. If snapshot-based, verify that snapshot $s$ is not de-authorized; if selective, verify that each target $(O_i, k_i)$ is not de-authorized under the system's per-version policy. Abort on failure.
  \item \textbf{Historical Version Authentication.} For each $(O_i, k_i) \in T$, verify inclusion proof in $\mathcal{V}$ under $R$, retrieving content digest $h_i^{k_i}$ from the entry at counter $k_i$.
  \item \textbf{Version Catalog Extension.} For each affected object append entry to $\mathcal{V}$ mapping $(O_i, c')$ to $h_i^{k_i}$ (reusing the historical digest) where $c' = c+1$ is the next counter value; update head pointer for $O_i$ to $c'$. Record $k_i$ as the \texttt{origin} pointer in the version metadata to link the new version to its historical source, enabling rollback detection in lineage reconstruction. Earlier entries remain immutable.
    \item \textbf{Completion Logging.} Append completion record to $\mathcal{L}$ linking $txid$, the intent, and the affected objects with their new counter-indexed versions.
    \item \textbf{Sealing.} Compute roots $H_{\mathcal{V}}$, $H_{\mathcal{S}}$, $H_{\mathcal{L}}$, aggregate into $R'$, and compute $\sigma' = \texttt{Seal}(R', c')$ where $c' = c+1$.
    \item \textbf{Publication.} Durably store $(R', c', \sigma')$, then advance the hardware monotonic counter from $c$ to $c'$ to finalize $(R,c) \leftarrow (R', c')$, making $R'$ authoritative.
\end{enumerate}
\end{mdframed}
\caption{Rollback protocol: verifies eligibility in $\mathcal{V}$ and $\mathcal{S}$ creates new version entries reusing historical digests, logs to $\mathcal{L}$, and co-seals under $R'$.}
\label{fig:rollback-protocol}
\end{figure}

\begin{figure}[!h]
\centering
\small
\begin{mdframed}[linewidth=1pt,linecolor=gray!30,backgroundcolor=gray!5]
	\textbf{Pruning Protocol.} \textit{Input:} current authoritative state $(R,c)$; pruning request specifying either snapshot tag $t$ (to de-authorize the snapshot composition) or explicit retire set $P = \{(O_i, k_i)\}$ where $k_i$ denotes historical counter values, plus reason/policy context (e.g., \texttt{retention\_expired}, CVE id, redaction tag) and administrator justification. \textit{Output:} new authoritative state $(R', c', \sigma')$ and authenticated log records.
\begin{enumerate}[leftmargin=1.1em, itemsep=2pt]
    \item \textbf{Intent Declaration.} Generate a fresh transaction identifier $txid$; append intent record to $\mathcal{L}$ with prior root $R$, mode (snapshot-based or selective), snapshot tag or explicit candidate set $P$ (object identifiers with counter values), policy context, actor, and justification.
  \item \textbf{Target Resolution.} If snapshot-based, treat the target as the snapshot tag $t$ itself (i.e., the composition recorded in $\mathcal{S}$); if selective, use the caller-provided retire set $P = \{(O_i, k_i)\}$.
  \item \textbf{Eligibility Validation.} Enforce retention and revocation policies. If snapshot-based, verify snapshot tag $t$ exists in $\mathcal{S}$ under $R$ and is not already de-authorized; if selective, for each $(O_i, k_i) \in P$ verify inclusion in $\mathcal{V}$ under $R$ and that no per-version de-authorization tombstone exists. Abort on failure.
  \item \textbf{De-Authorization Records.} If snapshot-based, append a tombstone record that marks snapshot tag $t$ ineligible for rollback (without revoking underlying object versions). If selective, append a tombstone entry for each validated $(O_i, k_i)$ to mark that version ineligible. Each tombstone includes metadata $\{reason, txid, justification, timestamp\}$.
  \item \textbf{Completion Logging.} Append completion record to $\mathcal{L}$ linking $txid$, the intent, and the finalized de-authorization decision.
    \item \textbf{Sealing.} Compute roots $H_{\mathcal{V}}$, $H_{\mathcal{S}}$, $H_{\mathcal{L}}$, aggregate into $R'$, and obtain $\sigma' = \texttt{Seal}(R', c')$ where $c' = c+1$.
    \item \textbf{Publication.} Durably store $(R', c', \sigma')$, then advance the hardware monotonic counter from $c$ to $c'$ to finalize $(R,c) \leftarrow (R', c')$, making $R'$ authoritative. Payload reclamation (deleting content from untrusted storage) may occur \emph{after} durable sealing and counter advance.
\end{enumerate}
\end{mdframed}
\caption{Pruning protocol: verifies pruning target(s) under $R$ (snapshot tag existence/eligibility in $\mathcal{S}$ for snapshot-based pruning, or version existence/eligibility in $\mathcal{V}$ for selective pruning), appends the corresponding tombstone record (tag tombstone in $\mathcal{S}$ or per-version tombstones in $\mathcal{V}$), logs to $\mathcal{L}$, and co-seals under $R'$.}
\label{fig:pruning-protocol}
\end{figure}

\clearpage
\begin{table*}[!h]
\centering
\scriptsize
\caption{Analytical costs per operation in \emph{snapshot-based mode}. $n$ denotes the number of objects updated in a batch; $n_s$ denotes the number of objects bound to a snapshot (i.e., the rollback set size); $S_{\mathcal{L}}$ denotes the number of leaves in the audit log $\mathcal{L}$. ``Leaves'' counts per-leaf appends to the PADs ($\mathcal{V}$, $\mathcal{S}$, $\mathcal{L}$, and head tracking). ``Proofs'' counts the number of inclusion/non-membership proof \emph{verifications}. ``Total Steps'' counts high-level protocol steps and iterations (e.g., per-object processing, scanning log entries), and is not a precise runtime complexity.}

\label{tab:analytical-costs}
\begin{tabular}{lcccccc}
	\toprule
	& \textbf{State update} & \textbf{Snapshot} & \textbf{Rollback to snapshot} & \textbf{Prune snapshot} & \textbf{Verify snapshot entry} & \textbf{Reconstruct lineage} \\
	\midrule
	\textbf{New Leaves} & $n{+}2$ & $3$ & $n_s{+}2$ & $3$ & $0$ & $0$ \\
	\textbf{Seal ops} & $1$ & $1$ & $1$ & $1$ & $0$ & $0$ \\
	\textbf{Counter incs} & $1$ & $1$ & $1$ & $1$ & $0$ & $0$ \\
  \textbf{\# Proofs} & $O(1)$ & $O(1)$ & $O(n_s)$ & $O(1)$ & $O(1)$ & $O(S_{\mathcal{L}})$ \\
  \textbf{Total Steps} & $O(n)$ & $O(n_s)$ & $O(n_s)$ & $O(1)$ & $O(1)$ & $O(S_{\mathcal{L}})$ \\
  \textbf{Total Storage} & $O(n)$ tuples + audit & $O(n_s)$ tag map + audit & $O(n_s)$ tuples + audit & $O(1)$ tombstone + audit & $0$ & $0$ \\
	\bottomrule
\end{tabular}
\end{table*}

\begin{table*}[!t]
  \centering
 \caption{Feature and performance comparison of \tool{} and prior state continuity frameworks. \tool{} provides several new security guarantees and capabilities. (* = Per-object update latency is higher because \tool{} uses a Merkle tree to enable efficient queries against history. Nimble uses a simple hash chain because it is not designed to query history, which would otherwise require a costly full linear scan.)}
  \label{tab:rebound_vs_nimble}
    \begin{tabular}{@{}lll@{}}
      \toprule
      \textbf{Feature} & \textbf{State Continuity Frameworks~\cite{angel2023nimble,niu2022narrator,parno2011memoir,strackx2016ariadne,matetic2017rote}} & \textbf{\tool{}} \\
      \midrule
      
      \textbf{Rollback Protection} & Yes & Yes \\
      \textbf{Versioning} & No (sealed latest-state digest) & Yes (whole-snapshot and per-object) \\
      \textbf{Authorized Rollback} & No (treated as adversarial) & Yes (a first-class, policy-driven operation) \\
      \textbf{Auditability} & Limited (cannot represent/audit legitimate rollbacks) & Fine-grained (cause, scope, and effect of all updates) \\

      \textbf{Core Data Structure} & Hash chain over sealed state digests & Merkle Tree (PAD) \\
      
      \textbf{State update latency (Nimble)~\cite{angel2023nimble}} & \textasciitilde 30--40ms per object & \textasciitilde 400-1000ms per object (from history length 0-100)* \\
      
      \bottomrule
    \end{tabular}%
\end{table*}

\clearpage
\section*{Concurrency Considerations}

\begin{figure}[!h]
    \centering
    \includegraphics[width=0.45\textwidth]{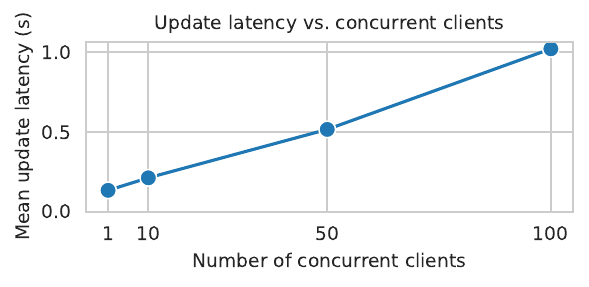}
    \includegraphics[width=0.45\textwidth]{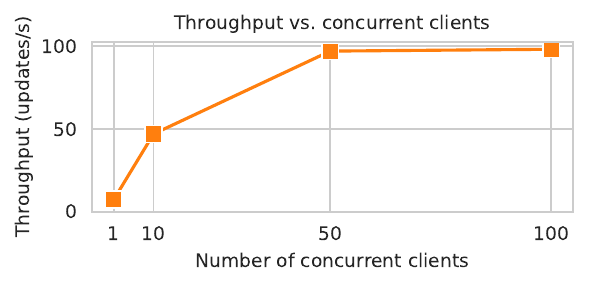}
    \caption{State update latencies increase linearly with respect to the number of concurrent clients updating the state, while throughput increases and then plateaus.}
    \label{fig:micro-concurrency}
\end{figure}

\red{Concurrency in CI/CD can arise when many repositories trigger release workflows at once (e.g., synchronized nightly builds, coordinated multi-service rollouts, or bursty merge queues after review cycles), causing overlapping update requests to the continuity layer. Figure~\ref{fig:micro-concurrency} shows that mean update latency increases approximately linearly with the number of concurrent clients, which is expected under increasing contention, while throughput rises and then plateaus, also as expected once the system approaches saturation. Importantly, observed update latencies remain sub-second even at high concurrency, and the throughput increase with concurrency indicates that the implementation benefits from batching to improve performance in concurrent settings.}

\section*{LLM usage considerations}
LLMs were used to assist with the Go implementation, primarily for generating unit tests and documentation for other developers. LLMs were also used for editorial purposes in this manuscript, and all outputs were inspected by the authors to ensure accuracy and originality. All ideas, technical content, experiments, etc., are original work of the authors.

%% file: meta_review.tex
\clearpage
\newpage 

\appendices 

\section{Meta-Review}

The following meta-review was prepared by the program committee for the 2026
IEEE Symposium on Security and Privacy (S\&P) as part of the review process as
detailed in the call for papers.

\subsection{Summary}
This paper proposes a novel system to provide state continuity support for confidential cloud applications. The main novelty lies in the fact that the proposed system enables policy-authorized rollbacks with full auditability. The paper also presents and evaluates an implementation of the system in Go and quantifies the performance and storage overhead.

\subsection{Scientific Contributions}
\begin{itemize}
\item Provides a Valuable Step Forward in an Established Field.
\end{itemize}

\subsection{Reasons for Acceptance}
\begin{enumerate}
\item While there is a significant body of work on the foundations of providing state continuity, the paper provides a valuable step forward in this field by identifying and solving an important practical problem: the need to balance the desire for rollback protection with the operational reality of bugs and incidents necessitating rollbacks.
\item The paper is well-presented. It provides a meticulous analysis of the problem of secure and auditable state rollback, with systematic descriptions of security properties and design goals.
\item The paper provides a pointer to the open-source implementation of the proposed system, thus supporting reproducibility of the results and enabling further work on improvements.
\end{enumerate}

\subsection{Noteworthy Concerns} 
\begin{enumerate} 
\item The paper focuses on enforcing authorization decisions, but leaves the challenging issue of how to specify authorization policies out of scope.
\end{enumerate}

